\documentclass[article]{jss}

\usepackage{amsbsy} 
\usepackage{amsmath}
\usepackage{amsfonts}
\usepackage{thumbpdf}
\usepackage{subfigure}
\usepackage{float}
\usepackage{tabularx}
\usepackage{graphicx}
\usepackage{epstopdf}
\usepackage[latin1]{inputenc}
\usepackage[usenames,dvipsnames]{xcolor}
\newcommand{\real}{\rm I\!R}

\author{Angelo  Mazza   \\University of Catania    \And 
        Antonio Punzo   \\University of Catania    \And 
        Brian   McGuire \\Montana State University
        }
\title{
\pkg{KernSmoothIRT}: An \proglang{R} Package for Kernel Smoothing in Item Response Theory
}

\Plainauthor{Angelo Mazza, Antonio Punzo, Brian McGuire} 
\Plaintitle{KernSmoothIRT: An R Package for Kernel Smoothing in Item Response Theory} 

\Abstract{
Item response theory (IRT) models are a class of statistical models used to describe the response behaviors of individuals to a set of items having a certain number of options. 
They are adopted by researchers in social science, particularly in the analysis of performance or attitudinal data, in psychology, education, medicine, marketing and other fields where the aim is to measure latent constructs.
Most IRT analyses use parametric models that rely on assumptions that often are not satisfied.
In such cases, a nonparametric approach might be preferable; nevertheless, there are not many software applications allowing to use that.  

To address this gap, this paper presents the \proglang{R} package \pkg{KernSmoothIRT}. 
It implements kernel smoothing for the estimation of option characteristic curves, and adds several plotting and analytical tools to evaluate the whole test/questionnaire, the items, and the subjects.
In order to show the package's capabilities, two real datasets are used, one employing multiple-choice responses, and the other scaled responses.
}

\Address{
  Angelo Mazza, Antonio Punzo\\
  Department of Economics and Business\\
  University of Catania\\
  Corso Italia, 55, 95129 Catania, Italy\\
  E-mail: \email{a.mazza@unict.it}, \email{antonio.punzo@unict.it}\\
  URL: \url{http://www.economia.unict.it/a.mazza}, \url{http://www.economia.unict.it/punzo}\\ \\
  Brian McGuire\\
  Department of Statistics\\
  Montana State University\\
  Bozeman, Montana, USA\\
  E-mail: \email{mcguirebc@gmail.com}
}

\begin{document}


\section[Introduction]{Introduction}
\label{sec:Introduction}


In psychometrics the analysis of the relation between latent continuous variables and observed dichotomous/polytomous variables is known as item response theory (IRT).
Observed variables arise from items of one of the following formats: \textit{multiple-choice} in which only one alternative is designed to be correct, \textit{multiple-response} in which more than one answer may be keyed as correct, \textit{rating scale} in which the phrasing of the response categories must reflect a scaling of the responses, \textit{partial credit} in which a partial credit is given in accordance with an examinee's degree of attainment in solving a problem, and \textit{nominal} in which there is neither a correct option nor an option ordering.     
Naturally, a set of items can be a mixture of these item formats.
Hereafter, for consistency's sake, the term ``option'' will be used as the unique term for several often used synonyms like: (response) category, alternative, answer, and so on; also the term ``test'' will be used to refer to a set of items comprising any psychometric test or questionnaire.

Our notation and framework can be summarized as follows.
Consider the responses of an $n$-dimensional set $\mathcal{S}=\left\{S_1,\ldots,S_i,\ldots,S_n\right\}$ of subjects to a $k$-dimensional sequence $\mathcal{I}=\left\{I_1,\ldots,I_j,\ldots,I_k\right\}$ of items.
Let $\mathcal{O}_j=\left\{O_{j1},\ldots,O_{jl},\ldots,O_{jm_j}\right\}$ be the $m_j$-dimensional set of options conceived for $I_j$, and let $x_{jl}$ be the weight attributed to $O_{jl}$.
The actual response of $S_i$ to $I_j$ can be so represented as a selection vector $\boldsymbol{y}_{ij}=\left(y_{ij1},\ldots,y_{ijm_j}\right)^\top$, where $\boldsymbol{y}_{ij}$ is an observation from the random variable $\boldsymbol{Y}_{ij}$ and $y_{ijl}=1$ if the option $O_{jl}$ is selected, and $0$ otherwise.
From now on it will be assumed that, for each item $I_j\in\mathcal{I}$, the subject selects one and only one of the $m_j$ options in $\mathcal{O}_j$; omitted responses are permitted.

The central problem in IRT, with reference to a generic option $O_{jl}$ of $I_j$, is the specification of a mathematical model describing the probability of selecting $O_{jl}$ as a function of $\vartheta$, the underlying latent trait which the test attempts to measure (the discussion is here restricted to models for items that measure one continuous latent variable, i.e., \textit{unidimensional latent trait models}).
According to \cite{Rams:kern:1991}, this function, or curve, will be referred to as \textit{option characteristic curve} (OCC), and it will be denoted with
\begin{equation*}
p_{jl}\left(\vartheta\right)=\Prob\left(\text{select $O_{jl}$}\left|\vartheta\right.\right)=\Prob\left(Y_{jl}=1\left|\vartheta\right.\right),                       
\end{equation*}
$j=1,\ldots,k$ and $l =1,\ldots,m_j$.
For example, in the analysis of multiple-choice items, which has typically relied on numerical statistics such as the proportion of subjects selecting each option and the point biserial correlation (quantifying item discrimination), it might be more informative to take into account all of the OCCs \citep[][]{Lei:Dunb:Kole:Acom:2004}.
Moreover, the OCCs are the starting point for a wide range of IRT analyses \citep[see, e.g.,][]{Bake:Kim:Item:2004}.
Note that, the term ``option characteristic curve'' is not by any means universal. 
Among the different terms found in literature, there are \textit{category characteristic curve}, \textit{operating characteristic curve}, \textit{category response function}, \textit{item category response function}, \textit{option response function} and more (see \citealp[][p.~10]{Osti:Neri:Poly:2006} and \citealp[][p.~23]{DeMa:Item:2010} for a survey of the different names used). 

With the aim to estimate the OCCs, in analogy with the classic statistical modeling, at least two routes are possible.
The first, and most common, is the \textit{parametric} one (PIRT: parametric IRT), in which a parametric structure is assumed so that the estimation of an OCC is reduced to the estimation of a vector parameter, of dimension varying from model to model, for each item in $\mathcal{I}$ (see, e.g., \citealp{This:Stei:taxo:1986}, \citealp{vand:Hamb:hand:1997}, \citealp{Osti:Neri:Poly:2006}, and \citealp{Neri:Osti:Hand:2010}, to have an idea of the existing PIRT models).
This vector is usually considered to be of direct interest and its estimate is often used as a summary statistic of some item aspects such as difficulty and discrimination \citep[see][]{Lord:appl:1980}.
The second route is the \textit{nonparametric} one (NIRT: nonparametric IRT), in which estimation is made directly on $\boldsymbol{y}_{ij}$, $i=1,\ldots,n$ and $j=1,\ldots,k$, without assuming any mathematical form for the OCCs, in order to obtain more flexible estimates which, according to \citet[][p.~348]{vand:Hamb:hand:1997}, can be assumed to be closer to the true OCCs than those provided by PIRT models. 
Accordingly, \cite{Rams:afun:1997} argues that NIRT might become the reference approach unless there are substantive reasons for preferring a certain parametric model.
Moreover, although nonparametric models are not characterized by parameters of direct interest, they encourage the graphical display of results; \citet[][p.~384]{Rams:afun:1997}, by personal experience, confirms the communication advantage of an appropriate display over numerical summaries.
These are only some of the motivations which justify the growth of NIRT research in recent years; other considerations can be found in \citet{Junk:Sijt:nonp:2001} who identify three broad motivations for the development and continued interest in NIRT. 

This paper focuses on NIRT.
Its origins -- prior to interest in PIRT -- are found in the scalogram analysis of \citet{Gutt:TheC:1947,Gutt:Rela:1950,Gutt:TheB:1950}.
Nevertheless, the work by \citet{Mokk:Athe:1971} is recognized as the first important contribution to this paradigm; he not only gave a nonparametric representation of the item characteristic curves in the form of a basic set of formal properties they should satisfy, but also provided the statistical theory needed to check whether these properties would hold in empirical data.
Among these properties, \textit{monotonicity} with respect to $\vartheta$ was required.
The \proglang{R} package \pkg{mokken} \citep{vanderArk:Mokken:2007} provides tools to perform a Mokken scale analysis.
Several other NIRT approaches have been proposed \citep[see][]{vand:rela:2001}.
Among them, kernel smoothing \citep{Rams:kern:1991} is a promising option, due to conceptual simplicity as well as advantageous practical and theoretical properties. 
The computer software \proglang{TestGraf} \citep{Rams:test:2000} performs kernel smoothing estimation of OCCs and related graphical analyses.
In this paper we present the \proglang{R} \citep{R} package \pkg{KernSmoothIRT}, available from CRAN (\url{http://CRAN.R-project.org/}), which offers most of the \proglang{TestGraf} features and adds some related functionalities.
Note that, although \proglang{R} is well-provided with PIRT techniques (see \citealt{deLe:Mair:Anin:2007} and \citealt{Wick:Stro:Zeil:Psych:2012}), it does not offer nonparametric analyses, of the type described above, in IRT. 
Nonparametric smoothing techniques of the kind found in \pkg{KernSmoothIRT} are commonly used and often cited exploratory statistical tools; as evidence, consider the number of times in which classical statistical studies use the functions \code{density()} and \code{ksmooth()}, both in the \pkg{stats} package, for kernel smoothing estimation of a density or regression function, respectively.
Consistent with its exploratory nature, \pkg{KernSmoothIRT} can be used as a complementary tool to other IRT packages; for example a \pkg{mokken} package user may use it to evaluate monotonicity.
OCCs smoothed by kernel techniques, due to their statistical properties (see \citealp{Doug:join:1997,Doug:asym:2001} and \citealp{Doug:Cohe:nonp:2001}), have been also used in PIRT analysis as a benchmark to estimate the best OCCs in a pre-specified parametric family \citep{Punz:Onke:2009}. 

The paper is organized as follows.
Section~\ref{sec:Kernel smoothing of OCCs} retraces kernel smoothing estimation of the OCCs and Section~\ref{sec:Functions related to the OCCs} illustrates other useful IRT functions based on these estimates.
The relevance of the package is shown, via two real data sets, in Section~\ref{sec:Package KernSmoothIRT in use}, and conclusions are finally given in Section~\ref{sec:conclusions}.
  
\section[Kernel smoothing of OCCs]{Kernel smoothing of OCCs}
\label{sec:Kernel smoothing of OCCs}

\citet{Rams:kern:1991,Rams:afun:1997} popularized nonparametric estimation of OCCs by proposing regression methods, based on kernel smoothing approaches, which are implemented in the \proglang{TestGraf} program \citep{Rams:test:2000}.
The basic idea of kernel smoothing is to obtain a nonparametric estimate of the OCC by taking a (local) weighted average (see \citealp{Altm:an:1992}, \citealp{Hard:Appl:1992}, and \citealp{Simo:smoo:1996}) of the form 
\begin{equation}
\widehat{p}_{jl}\left(\vartheta\right)=\sum_{i=1}^nw_{ij}\left(\vartheta\right)y_{ijl},     
\label{eq:NW kernel OCCs}
\end{equation}
$j=1,\ldots,k$ and $l=1,\ldots,m_j$, where the weights $w_{ij}\left(\vartheta\right)$ are defined so as to be maximal when $\vartheta=\vartheta_i$ and to be smoothly non-increasing as $\left|\vartheta-\vartheta_i\right|$ increases, with $\vartheta_i$ being the value of $\vartheta$ for $S_i\in\mathcal{S}$.
The need to keep $\widehat{p}_{jl}\left(\vartheta\right)\in\left[0,1\right]$, for each $\vartheta\in\real$, requires the additional constraints $w_{ij}\left(\vartheta\right)\geq 0$ and $\sum_{i=1}^nw_{ij}\left(\vartheta\right)=1$; as a consequence, it is preferable to use Nadaraya-Watson weights (\citealp{Nada:ones:1964} and \citealp{Wats:Smoo:1964}) of the form
\begin{equation}
w_{ij}\left(\vartheta\right)=\frac{\displaystyle K\left(\frac{\vartheta-\vartheta_i}{h_j}\right)}{\displaystyle\sum_{r=1}^nK\left(\frac{\vartheta-\vartheta_r}{h_j}\right)},       \label{eq:NW weights}
\end{equation}
where $h_j>0$ is the \textit{smoothing parameter} (also known as \textit{bandwidth}) controlling the amount of smoothness (in terms of bias-variance trade-off), while $K$ is the \textit{kernel function}, a nonnegative, continuous ($\widehat{p}_{jl}$ inherits the continuity from $K$) and usually symmetric function that is non-increasing as its argument moves further from zero.

Since the performance of \eqref{eq:NW kernel OCCs} largely depends on the choice of $h_j$, rather than on the kernel function \citep[see, e.g.,][]{Marr:Nola:Cano:1988} a simple Gaussian kernel $K\left(u\right)=\exp\left(-u^2/2\right)$ is often preferred (this is the only setting available in \proglang{TestGraf}).
Nevertheless, \pkg{KernSmoothIRT} allows for other common choices such as the uniform kernel $K\left(u\right)=\mathbb{I}_{\left[-1,1\right]}\left(u\right)$, and the quadratic kernel $K\left(u\right)=\left(1-u^2\right)\mathbb{I}_{\left[-1,1\right]}\left(u\right)$, where $\mathbb{I}_A\left(u\right)$ represents the indicator function assuming value 1 on $A$ and 0 otherwise.
In addition to the functionalities implemented in \proglang{TestGraf}, \pkg{KernSmoothIRT} allows the bandwidth $h_j$ to vary from item to item (as highlighted by subscript $j$).
This is an important aspect, since different items may not require the same amount of smoothing to obtain smooth curves \citep[][p.~8]{Lei:Dunb:Kole:Acom:2004}.

\subsection[Estimating abilities]{Estimating abilities}
\label{sec:Estimating abilities}

Unlike the standard kernel regression methods, in \eqref{eq:NW kernel OCCs} the dependent variable $Y_{jl}$ is a binary variable and the independent one is the latent variable $\vartheta$. 
Although $\vartheta$ cannot be directly observed, kernel smoothing can still be used, but each $\vartheta_i$ in \eqref{eq:NW weights} must be replaced with a reasonable estimate $\widehat{\vartheta}_i$ \citep{Rams:kern:1991} leading to  
\begin{equation}     
\widehat{p}_{jl}\left(\vartheta\right)=\displaystyle\sum_{i=1}^n\widehat{w}_i\left(\vartheta\right)y_{ijl},                   \label{eq:NWIRTordinal}
\end{equation}
where
\begin{displaymath}
\widehat{w}_i\left(\vartheta\right)=\frac{\displaystyle K\left(\frac{\vartheta-\widehat{\vartheta}_i}{h_j}\right)}{\displaystyle\sum_{r=1}^nK\left(\frac{\vartheta-\widehat{\vartheta}_r}{h_j}\right)}.	
\end{displaymath}

The choice of the scale of $\widehat{\vartheta}_i$ is arbitrary, since in this context only rank order considerations make sense (\citealp{Bart:Late:1983} and \citealp[][p.~614]{Rams:kern:1991}). 
Therefore, as most IRT models do, the estimation process begins (\citealp[][p.~615]{Rams:kern:1991} and \citealp[][pp.~25--26]{Rams:test:2000}) with:
\begin{enumerate}
	\item computation of the transformed rank $r_i=\mathsf{rank}\left(S_i\right)/\left(n+1\right)$, with $\mathsf{rank}\left(S_i\right)\in\left\{1,\ldots,n\right\}$, induced by some suitable statistic $t_i$, the total score 
\begin{displaymath}
t_i=\sum_{j=1}^{k}\sum_{l=1}^{m_j}y_{ijl}x_{jl}
\end{displaymath}
being the most obvious choice.
	\pkg{KernSmoothIRT} also allows, through the argument \code{RankFun} of the \code{ksIRT()} function, for the use of common summary statistics available in \proglang{R}, such as \code{mean()} and \code{median()}, or for a custom user-defined function. Alternatively, the user may specify the rank of each subject explicitly through the argument \code{SubRank}, allowing subject ranks to come from another source than the test being studied.
	\item replacement of $r_i$ by the quantile $\widehat{\vartheta}_i$ of some distribution function $F$.
	The estimated ability value for $S_i$ then becomes $\widehat{\vartheta}_i=F^{-1}\left(r_i\right)$.
	In these terms, the denominator $n+1$ of $r_i$ avoids an infinity value for the biggest $\widehat{\vartheta}_i$ when $\lim_{\vartheta\rightarrow + \infty }F\left(\vartheta\right)=1^-$.
Note that the choice of $F$ is equivalent to the choice of the $\vartheta$-metric. 
Historically, the standard Gaussian distribution $F=\Phi$ has been heavily used \citep[see][]{Bart:Thes:1988}.
  However, \pkg{KernSmoothIRT} allows the user specification of $F$ through one of the classical continuous distributions available in \proglang{R}.          
\end{enumerate}
Since these preliminary ability estimates are rank-based, they are usually referred to as \textit{ordinal ability estimates}.
Note that even a substantial amount of error in the ranks has only a small impact on the estimated curve values. 
This can be demonstrated both by mathematical analysis and through simulated data (see \citealp{Rams:kern:1991,Rams:test:2000} and \citealp{Doug:join:1997}).
Further theoretical results can be found in \citet{Doug:asym:2001} and \citet{Doug:Cohe:nonp:2001}.
The latter also assert that, if nonparametric estimated curves are meaningfully different from parametric ones, this parametric model -- defined on the particular scale determined by $F$ -- is an incorrect model for the data. 
In order to make this comparison valid, it is fundamental that the same $F$ is used for both nonparametric and parametric curves.
Thus, in the choice of a parametric family, visual inspections of the estimated kernel curves can be useful \citep{Punz:Onke:2009}.  

\subsection[Operational aspects]{Operational aspects}
\label{subsec:Operational aspects}
     
Operationally, the kernel OCC is evaluated on a finite grid, $\vartheta_1,\ldots,\vartheta_s,\ldots,\vartheta_q$, of $q$ equally-spaced values spanning the range of the ordinal ability estimates, so that the distance between two consecutive points is $\delta$.
Thus, starting from the values of $y_{ijl}$ and $\widehat{\vartheta}_i$, by grouping we can define the two sequences of $q$ values
\begin{displaymath}
\widetilde{y}_{sjl}=\sum_{i=1}^n\mathbb{I}_{\left[\vartheta_s-\delta/2,\vartheta_s+\delta/2\right)}\left(\widehat{\vartheta}_i\right)y_{ijl}\quad \text{and} \quad v_s=\sum_{i=1}^n\mathbb{I}_{\left[\vartheta_s-\delta/2,\vartheta_s+\delta/2\right)}\left(\widehat{\vartheta}_i\right).	
\end{displaymath}
Up to a scale factor, the sequence $\widetilde{y}_{sjl}$ is a grouped version of $y_{ijl}$, while $v_s$ is the corresponding number of subjects in that group.
It follows that
\begin{equation}     
\widehat{p}_{jl}\left(\vartheta\right)\approx\displaystyle\frac{\displaystyle\sum_{s=1}^q K\left(\frac{\vartheta-\vartheta_s}{h_j}\right)\widetilde{y}_{sjl}}{\displaystyle\sum_{s=1}^qK\left(\frac{\vartheta-\vartheta_s}{h_j}\right)v_s},\qquad \vartheta\in \Bigl\{\vartheta_1,\ldots,\vartheta_s,\ldots,\vartheta_q\Bigr\}.                   \label{eq:practical kernel estimates}
\end{equation}

\subsection[Choosing the smoothing parameter]{Cross-validation selection for the bandwidth}
\label{subsec:cross-validation}

Two of the most frequently used methods of bandwidth selection are the plug-in method and the cross-validation \citep[for a more complete treatment of these methods see, e.g.,][]{Hard:Appl:1992}. 

The former approach, widely used in kernel density estimation, often leads to rules of thumb.
Motivated by the need to have fast automatically generated kernel estimates, the function \code{ksIRT()} of \pkg{KernSmoothIRT} adopts, as default, the common rule of thumb of \citet[][p.~45]{Silv:dens:1986} for the Gaussian kernel density estimator.
It, in our context, is formulated as 
\begin{equation}
h_j=h=1.06\,\sigma_{\vartheta}\,n^{-1/5},
\label{eq:rule of thumb}
\end{equation}  
where $\sigma_{\vartheta}$ -- that in the original framework is a sample estimate -- simply represents the standard deviation of $\vartheta$, induced by $F$.
Note that \eqref{eq:rule of thumb}, with $\sigma_{\vartheta}=1$, is the unique approach considered in \texttt{TestGraf}.  

The second approach, cross-validation, requires a considerably higher computational effort; nevertheless, it is simple to understand and widely applied in nonparametric kernel regression (see, e.g., \citealp{Wong:Onth:1983}, \citealp{Rice:Band:1984} and \citealp{Mazz:Punz:Disc:2011,Mazz:Punz:Grad:2012,Mazz:Punz:Usin:2013,Mazz:Punz:DBKG:2014}).
Its description, in our context, is as follows. 
Let $\boldsymbol{y}_j=\left(\boldsymbol{y}_{1j},\ldots,\boldsymbol{y}_{ij},\ldots,\boldsymbol{y}_{nj}\right)$ be the $m_j\times n$ selection matrix referred to $I_j$.
Moreover, let
\begin{displaymath}
\widehat{\boldsymbol{p}}_j\left(\vartheta\right)=\bigl(\widehat{p}_{j1}\left(\vartheta\right),\ldots,\widehat{p}_{jm_j}\left(\vartheta\right)\bigr)^\top	
\end{displaymath}
be the $m_j$-dimensional vector of kernel-estimated probabilities, for $I_j$, at the evaluation point $\vartheta$.
The probability kernel estimate evaluated in $\vartheta$, for $I_i$, can thus be written as
\begin{displaymath}
\widehat{\boldsymbol{p}}_j\left(\vartheta\right)=\sum_{i=1}^n\widehat{w}_{ij}\left(\vartheta\right)\boldsymbol{y}_{ij}=\boldsymbol{y}_j\widehat{\boldsymbol{w}}_j\left(\vartheta\right),  
\end{displaymath}  
where $\widehat{\boldsymbol{w}}_j\left(\vartheta\right)=\bigl(\widehat{w}_{1j}\left(\vartheta\right),\ldots,\widehat{w}_{ij}\left(\vartheta\right),\ldots,\widehat{w}_{nj}\left(\vartheta\right)\bigr)^\top$ denotes the vector of weights.

In detail, cross-validation simultaneously fits and smooths the data contained in $\boldsymbol{y}_j$ by removing one ``data point'' $\boldsymbol{y}_{ij}$ at a time, estimating the value of $\boldsymbol{p}_j$ at the correspondent ordinal ability estimate $\widehat{\vartheta}_i$, and then comparing the estimate to the omitted, observed value.
So the cross-validation statistic is
\begin{equation*}
\mathsf{CV}\left(h_j\right)=\frac{1}{n}\sum_{i=1}^n\biggl(\boldsymbol{y}_{ij}-\widehat{\boldsymbol{p}}_j^{\left(-i\right)}\left(\widehat{\vartheta}_i\right)\biggr)^\top\biggl(\boldsymbol{y}_{ij}-\widehat{\boldsymbol{p}}_j^{\left(-i\right)}\left(\widehat{\vartheta}_i\right)\biggr),
\end{equation*}
where 
\begin{displaymath}
\widehat{\boldsymbol{p}}_j^{\left(-i\right)}\left(\widehat{\vartheta}_i\right)=\frac{\displaystyle\sum_{\substack{r=1\\
r\neq i}}^n\displaystyle K\left(\frac{\widehat{\vartheta}_i-\widehat{\vartheta}_r}{h_j}\right)\boldsymbol{y}_{rj}}{\displaystyle\sum_{\substack{r=1\\
r\neq i}}^nK\left(\frac{\widehat{\vartheta}_i-\widehat{\vartheta}_r}{h_j}\right)}	
\end{displaymath}
is the estimated vector of probabilities at $\widehat{\vartheta}_i$ computed by removing the observed selection vector $\boldsymbol{y}_{ij}$, as denoted by the superscript in $\widehat{\boldsymbol{p}}_j^{\left(-i\right)}$.
The value of $h_j$ that minimizes $\mathsf{CV}\left(h_j\right)$ is referred to as the cross-validation smoothing parameter, $h_j^{\text{CV}}$, and it is possible to find it by systematically searching across a suitable smoothing parameter region.
 
\subsection[Pointwise confidence intervals]{Approximate pointwise confidence intervals}
\label{subsec:pointwise confidence intervals}

In visual inspection and graphical interpretation of the estimated kernel curves, pointwise confidence intervals at the evaluation points provide relevant information, because they indicate the extent to which the kernel OCCs are well defined across the range of $\vartheta$ considered.
Moreover, they are useful when nonparametric and parametric models are compared.

Since $\widehat{p}_{jl}\left(\vartheta\right)$ is a linear function of the data, as can be easily seen from \eqref{eq:NWIRTordinal}, and being $Y_{ijl}\sim\text{Ber}\left[p_{jl}\left(\widehat{\vartheta}_i\right)\right]$, 
\begin{eqnarray*}
\VAR\left[\widehat{p}_{jl}\left(\vartheta\right)\right]&=&\displaystyle\sum_{i=1}^n\left[\widehat{w}_i\left(\vartheta\right)\right]^2 \VAR\left(Y_{ijl}\right)\\
&=&\displaystyle\sum_{i=1}^n\left[\widehat{w}_i\left(\vartheta\right)\right]^2 p_{jl}\left(\widehat{\vartheta}_i\right)\left[1-p_{jl}\left(\widehat{\vartheta}_i\right)\right]. 
\end{eqnarray*}      
The above formula holds if independence of the $Y_{ijl}$s, with respect to the subjects, is assumed and possible error variation in the arguments, $\widehat{\vartheta}_i$, are ignored \citep{Rams:kern:1991}.
Substituting $p_{jl}$ for $\widehat{p}_{jl}$ yields the $\left(1-\alpha\right) 100\%$ approximate pointwise confidence intervals 
\begin{equation}
\widehat{p}_{jl}\left(\vartheta\right)\mp z_{1-\frac{\alpha}{2}}\sqrt{ \sum_{i=1}^n\left[\widehat{w}_i\left(\vartheta\right)\right]^2 \widehat{p}_{jl}\left(\widehat{\vartheta}_i\right)\left[1-\widehat{p}_{jl}\left(\widehat{\vartheta}_i\right)\right]}, \label{eq:confidence bands for ICRF}
\end{equation} 
where $z_{1-\frac{\alpha}{2}}$ is such that $\Phi\left[z_{1-\frac{\alpha}{2}}\right]=1-\frac{\alpha}{2}$.
Other more complicated approaches to interval estimation for kernel-based nonparametric regression functions are described in \citet{Azza:Bowm:Hard:Onth:1989} and \citet[][Section~4.2]{Hard:Appl:1992}.

\section[Functions related to the OCCs]{Functions related to the OCCs}
\label{sec:Functions related to the OCCs}

Once the kernel estimates of the OCCs are obtained, several other quantities can be computed based on them.
In what follows we will give a concise list of the most important ones.

\subsection[Expected item score]{Expected item score}
\label{subsec:Expected item score}

In order to obtain a single function for each item in $\mathcal{I}$ it is possible to define the expected value of the score $X_j=\sum_{l=1}^{m_j}x_{jl}Y_{jl}$, conditional on a given value of $\vartheta$ \citep[see, e.g.,][]{Chan:Mazz:uniq:1994}, as follows
\begin{equation}
e_j\left(\vartheta\right)=\E\left(X_j\left|\vartheta\right.\right)=\sum_{l=1}^{m_j}x_{jl}p_{jl}\left(\vartheta\right),                        \label{eq:EIS}
\end{equation} 
$j=1,\ldots,k$, that takes values in $\left[x_{j\min},x_{j\max}\right]$, where $x_{j\min}=\min\left\{x_{j1},\ldots,x_{jm_j}\right\}$ and $x_{j\max}=\max\left\{x_{j1},\ldots,x_{jm_j}\right\}$. 
The function $e_j\left(\vartheta\right)$ is commonly known as \textit{expected item score} (EIS) and can be viewed \citep{Lord:appl:1980} as a regression of the item score $X_j$ onto the $\vartheta$ scale.
Naturally, for dichotomous and multiple-choice IRT models, the EIS coincides with the OCC referred to the correct option.

Starting from \eqref{eq:EIS}, it is straightforward to define the kernel EIS estimate as follows  
\begin{equation*}
\widehat{e}_j\left(\vartheta\right)=\sum_{l=1}^{m_j}x_{jl}\widehat{p}_{jl}\left(\vartheta\right)=\sum_{l=1}^{m_j}x_{jl}\sum_{i=1}^n\widehat{w}_{ij}\left(\vartheta\right) y_{ijl}=\sum_{i=1}^n\widehat{w}_{ij}\left(\vartheta\right)\sum_{l=1}^{m_j}x_{jl}y_{ijl}.    
\end{equation*} 
For the EIS, in analogy with Section~\ref{subsec:pointwise confidence intervals}, the $\left(1-\alpha\right) 100\%$ approximate pointwise confidence interval is given by
\begin{equation}
\widehat{e}_j\left(\vartheta\right)\mp z_{1-\frac{\alpha}{2}}\sqrt{\widehat{\VAR\left[\widehat{e}_j\left(\vartheta\right)\right]}},  \label{eq:confidence bands for IRF}               
\end{equation}
and, since $Y_{ijl}Y_{ijt}\equiv 0$ for $l\neq t$, one has
\begin{displaymath}
\VAR\left[\widehat{e}_j\left(\vartheta\right)\right]=\displaystyle\sum_{i=1}^n\left[\widehat{w}_{ij}\left(\vartheta\right)\right]^2\VAR\left(\sum_{l=1}^{m_j}x_{jl}Y_{ijl}\right)
\end{displaymath}
where
\begin{eqnarray*}
\VAR\left(\sum_{l=1}^{m_j}x_{jl}Y_{ijl}\right)&=&\sum_{l=1}^{m_j}x_{jl}^2\VAR\left(Y_{ijl}\right)+\displaystyle\sum_{l=1}^{m_j}\sum_{t\neq l}x_{jl}x_{jt}\COV\left(Y_{ijl},Y_{ijt}\right)\\
&=&\sum_{l=1}^{m_j}x_{jl}^2\VAR\left(Y_{ijl}\right)-\displaystyle\sum_{l=1}^{m_j}\sum_{t\neq l}x_{jl}x_{jt}\E\left(Y_{ijl}\right)\E\left(Y_{ijt}\right)\\
&=&\sum_{l=1}^{m_j}x_{jl}^2p_{jl}\left(\widehat{\vartheta}_i\right)\left[1-p_{jl}\left(\widehat{\vartheta}_i\right)\right]-\displaystyle\sum_{l=1}^{m_j}\sum_{t\neq l}x_{jl}x_{jt}p_{jl}\left(\widehat{\vartheta}_i\right)p_{jt}\left(\widehat{\vartheta}_i\right).
\end{eqnarray*}
Substituting $p_{jl}$ with $\widehat{p}_{jl}$ in $\VAR\left[\widehat{e}_i\left(\vartheta\right)\right]$, one obtains $\widehat{\VAR\left[\widehat{e}_i\left(\vartheta\right)\right]}$, quantity that has to be inserted in \eqref{eq:confidence bands for IRF}.

Really, intervals in \eqref{eq:confidence bands for ICRF} and \eqref{eq:confidence bands for IRF} are, respectively, intervals for $\E\left[\widehat{p}_{jl}\left(\vartheta\right)\right]$ and $\E\left[\widehat{e}_j\left(\vartheta\right)\right]$, rather than for $p_{jl}\left(\vartheta\right)$ and $e_j\left(\vartheta\right)$; thus, they share the bias present in $\widehat{p}_{jl}$ and $\widehat{e}_j$, respectively \citep[for the OCC case, see][p.~619]{Rams:kern:1991}.

\subsection[Expected test score]{Expected test score}
\label{subsec:Expected test score}

In analogy to Section~\ref{subsec:Expected item score}, a single function for the whole test can be obtained as follows
\begin{equation*}
e\left(\vartheta\right)=\sum_{j=1}^{k}e_j\left(\vartheta\right)=\sum_{j=1}^{k}\sum_{l=1}^{m_j}x_{jl}p_{jl}\left(\vartheta\right).                        
\end{equation*} 
It is called \textit{expected test score} (ETS).
Its kernel smoothed counterpart can be specified as
\begin{equation}
\widehat{e}\left(\vartheta\right)=\sum_{j=1}^{k}\sum_{l=1}^{m_j}x_{jl}\widehat{p}_{jl}\left(\vartheta\right)
\label{eq:expected total score}
\end{equation} 
and may be preferred in substitution of $\vartheta$, for people who are not used to IRT, as display variable on the $x$-axis to facilitate the interpretation of the OCCs, as well as of other output-plots of \pkg{KernSmoothIRT}.
This possibility is considered through the argument \code{axistype="scores"} of the \code{plot()} method.
Note that, although it can happen that \eqref{eq:expected total score} fails to be completely increasing in $\vartheta$, this event is rare and tends to affect the plots only at extreme trait levels.

\subsection[Relative credibility curve]{Relative credibility curve}
\label{subsec:Relative Credibility Curve}

For a generic subject $S_i\in\mathcal{S}$, we can compute the relative likelihood 
\begin{equation}
L_i\left(\vartheta\right)=M^{-1}\displaystyle\prod_{j=1}^k\prod_{l=1}^{m_j}\left[\widehat{p}_{jl}\left(\vartheta\right)\right]^{y_{ijl}}
\label{eq:relative likelihood}
\end{equation}
of the various values of $\vartheta$ given his pattern of responses and given the kernel-estimated OCCs.
In \eqref{eq:relative likelihood}, $M=\max_{\vartheta }\left\{\prod_{j=1}^k\prod_{l=1}^{m_j}\left[\widehat{p}_{jl}\left(\vartheta\right)\right]^{y_{ijl}}\right\}$.
The function in \eqref{eq:relative likelihood} is also known as \textit{relative credibility curve} \citep[RCC; see, e.g,][]{Lind:Infe:1973}.
The $\vartheta$-value, say $\widehat{\vartheta}^{\text{ML}}$, such that $L_i\left(\vartheta\right)=1$, is called the maximum likelihood (ML) estimate of the ability for $S_i$ \citep[see also][]{Kuty:nonp:1997}.
Differently from simple summary statistics like the total score, $\widehat{\vartheta}^{\text{ML}}$ considers, in addition to the whole pattern of responses, also the characteristics of the items as described by their OCCs; thus, it will tend to be a more accurate estimate of the ability.

Finally, as \citet{Kuty:nonp:1997} and \citet{Rams:test:2000} do, the obtained values of $\widehat{\vartheta}^{\text{ML}}$ may be used as a basis for a second step of the kernel smoothing estimation of OCCs.
This iterative process, consisting in cycling back the values of $\widehat{\vartheta}^{\text{ML}}$ into estimation, can clearly be repeated any number of times with the hope that each step refines or improves the estimates of $\vartheta$.
However, as \citet{Rams:test:2000} states, for the vast majority of applications, no iterative refinement is really necessary, and the use of $\widehat{\vartheta}_i$ or $\widehat{\vartheta}^{\text{ML}}_i$ for ranking examinees works fine. 
This is the reason why we have not considered the iterative process in the package. 

\subsection[Probability simplex]{Probability simplex}
\label{subsec:Probability Simplex}

With reference to a generic item $I_j\in\mathcal{I}$, the vector of probabilities $\widehat{\boldsymbol{p}}_{j}\left(\vartheta\right)$ can be seen as a point in the probability simplex $\mathbb{S}^{m_j}$, defined as the $\left(m_j-1\right)$-dimensional subset of the $m_j$-dimensional space containing vectors with nonnegative coordinates summing to one.
As $\vartheta$ varies, because of the assumptions of smoothness and unidimenionality of the latent trait, $\widehat{\boldsymbol{p}}_{j}\left(\vartheta\right)$ moves along a curve; the item analysis problem is to locate the curve properly within the simplex.
On the other hand, the estimation problem for $S_i$ is the location of its position along this curve.

\begin{figure}[!ht]
\centering
\subfigure[Triangle \label{fig:triangle}]
{\resizebox{0.50\textwidth}{!}{\includegraphics{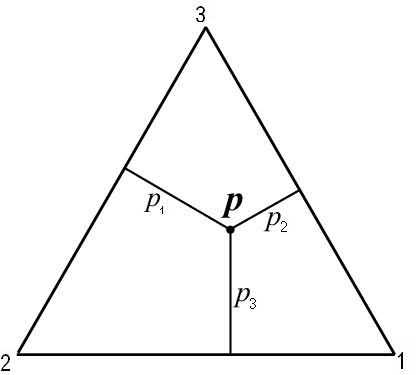}}}
\quad
\subfigure[Tetrahedron \label{fig:tetrahedron}]
{\resizebox{0.44\textwidth}{!}{\includegraphics{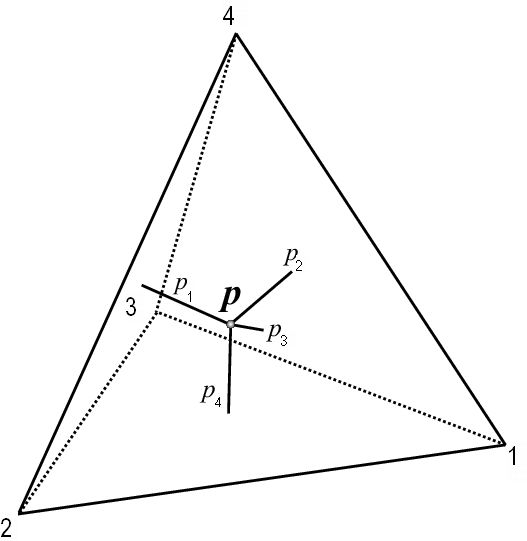}}}
\caption{Convenient way of displaying a point in the probability simplex $\mathbb{S}^{m_j}$ when $m_j=3$ (on the left) and $m_j=4$ (on the right).
\label{fig:probability simplex}
}
\end{figure}

As illustrated in \citet[][pp.~5--9]{Aitc:stat:2003}, a convenient way of displaying points in $\mathbb{S}^{m_j}$, when $m_j=3$ or $m_j=4$, is represented, respectively, by the \textit{reference triangle} in \figurename~\ref{fig:triangle} -- an equilateral triangle having unit altitude -- and by the \textit{regular tetrahedron}, of unit altitude, in \figurename~\ref{fig:tetrahedron}.
Here, for any point $\boldsymbol{p}$, the lengths of the perpendiculars $p_1,\ldots,p_{m_j}$ from $\boldsymbol{p}$ to the sides opposite to the vertices $1,\ldots,m_j$ are all greater than, or equal to, zero and have a unitary sum.
Since there is a unique point with these perpendicular values, there is a one-to-one correspondence between $\mathbb{S}^3$ and points in the reference triangle, and between $\mathbb{S}^4$ and points in the regular tetrahedron. 
Thus, we have a simple means for representing the vector of probabilities $\widehat{\boldsymbol{p}}_{j}\left(\vartheta\right)$ when $m_j=3$ and $m_j=4$.
Note that for items with more than four options there is no satisfactory way of obtaining a visual representation of the corresponding probability simplex; nevertheless, with \pkg{KernSmoothIRT} we can perform a partial analysis which focuses only on three or four of the options.

\section[Package description and illustrative examples]{Package description and illustrative examples}
\label{sec:Package KernSmoothIRT in use}

The main function of the package is \code{ksIRT()}; it creates an S3 object of class \code{ksIRT}, which provides a \code{plot()} method as well as a suite of functions that allow the user to analyze the subjects, the options, the items, and the overall test.
What follows is an illustration of the main capabilities of the \pkg{KernSmoothIRT} package.

\subsection[Kernel smoothing with the ksIRT function]{Kernel smoothing with the \code{ksIRT()} function}
\label{subsec:Kernel smoothing with the ksIRT function}

The \code{ksIRT()} function performs the kernel smoothing.
It requires \code{responses}, a $\left(n\times k\right)$-matrix, with a row for each subject in $\mathcal{S}$ and a column for each item in $\mathcal{I}$, containing the selected option numbers. 
Alternatively, \code{responses} may be a data frame or a list object.

\subsubsection[arguments to set the item format (format, key, weights)]{Arguments for setting the item format: \code{format}, \code{key}, \code{weights}}
\label{subsubsec:arguments to set the item format (format, key, weights)}

To use the basic weighting schemes associated with each item format, the following combination of arguments have to be applied.
\begin{itemize}
	\item For \textit{multiple-choice} items, use \code{format=1} and provide in \code{key}, for each item, the option number that corresponds to the correct option. 
	For \textit{multiple-response} items, one way to score them is simply to count the correctly classified options; to do this, a preliminary conversion of every option into a separate true/false item is necessary.
	\item For \textit{rating scale} and \textit{partial credit} items, use \code{format=2} and provide in \code{key} a vector with the number of options of every item.
	If all the items have the same number of options, then \code{key} may be a scalar. 
	\item For \textit{nominal} items, use \code{format=3}; \code{key} is omitted. 
	Note that to analyze items or options, subjects have to be ranked; this can only be done if the test also contains non-nominal items or if a prior ranking of subjects is provided with \code{SubRank}.
\end{itemize}
If the test is made of a mixture of different item formats, then \code{format} must be a numeric vector of length equal to the number of items.
More complicated weighting schemes may be specified using \code{weights} in lieu of both \code{format} and \code{key} (see the help for details).

\subsubsection[arguments for kernel smoothing (theta, nval, thetadist, kernel, bandwidth)]{Arguments for smoothing: \code{evalpoints}, \code{nevalpoints}, \code{thetadist}, \code{kernel}, \code{bandwidth}}
\label{subsubsec:arguments for kernel smoothing (theta, nval, thetadist, kernel, bandwidth)}

The user can select the $q$ evaluation points of Section~\ref{subsec:Operational aspects}, the ranking distribution $F$ of Section~\ref{sec:Estimating abilities}, the type of kernel function $K$ and the bandwidth(s).
The number $q$ of OCCs evaluation points may be specified in \code{nevalpoints}. 
By default they are 51 and their range is data dependent. 
Alternatively, the user may directly provide evaluation points using \code{evalpoints}.
As to $F$, it is by default $\Phi$; any other distribution, with its parameters values, may be provided in \code{thetadist}.
The default kernel function is the Gaussian; uniform or quadratic kernels may be selected with \code{kernel}. 
The global bandwidth is computed by default according to the rule of thumb in equation \eqref{eq:rule of thumb}. 
Otherwise, the user may either input a numerical vector of bandwidths for each item or opt for cross-validation estimation, as described in Section~\ref{subsec:cross-validation}, by specifying \code{bandwidth="CV"}.
  
\subsubsection[arguments to handle missing values (miss, NAweight)]{Arguments to handle missing values: \code{miss}, \code{NAweight}}
\label{subsubsec:arguments to handle missing values (miss, NAweight)}

Several approaches are implemented for handling missing answers.
The default, \code{miss="option"}, treats them as further options, with weight specified in \code{NAweight}, that by default is 0. 
When OCCs are plotted, the new options will be added to the corresponding items.
Other choices impute the missing values according to some discrete probability distributions taking values on $\left\{1,\ldots,m_j\right\}$, $j=1,\ldots,k$; the uniform distribution is specified by \code{miss="random.unif"}, while the multinomial distribution, with probabilities equal to the frequencies of the non-missing options of that item, is specified by \code{miss="random.multinom"}.
Finally, \code{miss="omit"} deletes from the data all the subjects with at least one omitted answer.

\subsubsection[the ksIRT class]{The \code{ksIRT} class}
\label{subsubsec:the ksIRT class }

The \code{ksIRT()} function returns an S3 object of class \code{ksIRT}; its main components, along with their brief descriptions, can be found in Table~\ref{tab:value}. 
Methods implemented for this class are illustrated in Table~\ref{tab:methods}.
The \code{plot()} method allows for a variety of exploratory plots, which are selected with the argument \code{plottype}; 
its main options are described in Table~\ref{tab:plottype}.
\begin{table}[!ht]
\centering
\begin{tabularx}{\linewidth}{l  X}
\hline
Values & Description \\ 
\hline
\code{\$itemcor}          & vector of item point polyserial correlations \\[2mm]
\code{\$evalpoints}       & vector of evaluation points used in curve estimation \\[2mm]
\code{\$subjscore}        & vector of observed subjects' overall scores \\[2mm]
\code{\$subjtheta}        & vector of subjects' quantiles on the distribution specified in \code{thetadist} \\[2mm]
\code{\$subjthetaML}      & vector of $\widehat{\vartheta}^{\text{ML}}_i$, $i=1,\ldots,n$ \\[2mm]
\code{\$subjscoreML}      & vector of subjects' ML scores $\widehat{e}\left(\widehat{\vartheta}^{\text{ML}}_i\right)$, $i=1,\ldots,n$ \\[2mm]
\code{\$subjscoresummary} & vector of quantiles, of probability 0.05, 0.25, 0.50, 0.75, 0.95, for the observed overall scores\\[2mm]
\code{\$subjthetasummary} & vector as \code{subjscoresummary} but computed on \code{subjtheta} \\[2mm]
\code{\$OCC}              & matrix of dimension $\left(\sum_{j=1}^km_j\right) \times \left(3+q\right)$.
The first three columns specify the item, the option, and the corresponding weight $x_{jl}$.
The additional columns contain the kernel smoothed OCCs at each evaluation point\\[2mm]
\code{\$stderrs}  & matrix as \code{OCC} containing the standard errors of \code{OCC}\\[2mm]
\code{\$bandwidth} & vector of bandwidths $h_j$, $j=1,\ldots,k$\\[2mm]
\code{\$RCC} & list of $n$ vectors containing the $q$ values of $L_i\left(\vartheta\right)$, $i=1,\ldots,n$\\ 
\hline
\end{tabularx}
\caption{List of most the important components of class \code{ksIRT}}
\label{tab:value}
\end{table}
\begin{table}[!ht]
\centering
\begin{tabularx}{\linewidth}{l  X}
\hline
Methods & Description \\ 
\hline
\code{plot()} & Allows for a variety of exploratory plots\\[2mm]
\code{itemcor()} & Returns a vector of item point polyserial correlations\\[2mm]
\code{subjscore()}   & Returns a vector of subjects' observed overall scores\\[2mm]
\code{subjthetaML()} & Returns a vector of $\widehat{\vartheta}^{\text{ML}}_i$, $i=1,\ldots,n$\\[2mm]
\code{subjscoreML()} & Returns a vector of $\widehat{e}\left(\widehat{\vartheta}^{\text{ML}}_i\right)$, $i=1,\ldots,n$\\[2mm]
\code{subjOCC()} & Returns a list of $k$ matrices, of dimension $\left(m_j\times n\right)$, $j=1,\ldots,k$, containing $\Prob\left(Y_{jl}=1\left|S_i\right.\right)$, $i=1,\ldots,n$ and $l=1,\ldots,m_j$.
The argument \code{stype} governs the scale on which to evaluate each subject; among the possible alternatives there are the observed score $t_i$ and the ML estimates $\widehat{\vartheta}_i^{\text{ML}}$, $i=1,\ldots,n$\\[2mm]
\code{subjEIS()} & Returns a $\left(k \times n\right)$ matrix of subjects' expected item scores\\[2mm]
\code{subjETS()} & Returns a vector of subjects' expected test scores\\[2mm]
\code{PCA()} & Returns a list of class \code{prcomp} of the \pkg{stats} package\\[2mm]
\code{subjOCCDIF()} & Returns a list containing, for each group, the same object returned by \code{subjOCC()}\\[2mm]
\code{subjEISDIF()} & Returns a list containing, for each group, the same object returned by \code{subjEIS()}\\[2mm]
\code{subjETSDIF()} & Returns a list containing, for each group, the same object returned by \code{subjETS()}\\
\hline
\end{tabularx}
\caption{Methods implemented in class \code{ksIRT}}
\label{tab:methods}
\end{table}
\begin{table}[!ht]
\centering
\begin{tabularx}{\linewidth}{l  X}
\hline
Option & Description \\ 
\hline
\code{"OCC"} & Plots the OCCs\\[2mm] 
\code{"EIS"} & Plots and returns the EISs\\[2mm] 
\code{"ETS"} & Plots and returns the ETS\\[2mm]
\code{"RCC"} & Plots the RCCs\\[2mm] 
\code{"triangle"/"tetrahedron"} & Displays a simplex plot with the highest 3 or 4 probability options\\[2mm]
\code{"PCA"} & Displays a PCA plot of the EISs\\[2mm]
\code{"OCCDIF"} & Plots OCCs for multiple groups\\[2mm]
\code{"EISDIF"} & Plots EISs for multiple groups\\[2mm] 
\code{"ETSDIF"} & Plots ETSs for multiple groups\\
\hline
\end{tabularx}
\caption{Main options for argument \code{plottype} of the \code{plot()} method.}
\label{tab:plottype}
\end{table}

\subsection[Psych 101]{Psych 101}
\label{subsec:Psych 101}

The first tutorial uses the Psych 101 dataset included in the \pkg{KernSmoothIRT} package. 
This dataset contains the responses of $n=379$ students, in an introductory psychology course, to $k=100$ multiple-choice items, each with $m_j=4$ options as well as a key. 
These data were also analyzed in \citet{Rams:Abra:Bino:1989} and in \citet{Rams:kern:1991}.

To begin the analysis, create a \code{ksIRT} object. 
This step performs the kernel smoothing and prepares the object for analysis using the many types of plots available.
\begin{CodeChunk}
\begin{CodeInput}
R> data("Psych101")
R> Psych1 <- ksIRT(responses=Psychresponses, key=Psychkey, format=1)
R> Psych1
\end{CodeInput}
\begin{CodeOutput}
    Item Correlation
1      1  0.23092838
2      2  0.09951663
3      3  0.19214764
.      .      .
.      .      .
.      .      .
99    99  0.01578162
100  100  0.24602614 
\end{CodeOutput}
\end{CodeChunk}
The command \code{data("Psych101")} loads both \code{Psychresponses} and \code{Psychkey}.    
The function \code{ksIRT()} produces kernel smoothing estimates using, by default, a Gaussian distribution $F$ (\code{thetadist=list("norm",0,1)}), a Gaussian kernel function $K$ (\code{kernel="gaussian"}), and the rule of thumb \eqref{eq:rule of thumb} for the global bandwidth.
The last command, \code{Psych1}, prints the point polyserial correlations, a traditional descriptive measure of item performance given by the correlation between each dichotomous/polythomous item and the total score \citep[see][for details]{Olss:Dras:Dora:Thep:1982}. 
As documented in Table~\ref{tab:methods}, these values can be also obtained via the command \code{itemcor(Psych1)}. 

Once the \code{ksIRT} object \code{Psych1} is created, several plots become available for analyzing each item, each subject and the overall test.  
They are displayed through the \code{plot()} method, as described below.
%

\subsubsection[Option Characteristic Curves]{Option characteristic curves}
\label{subsubsec:Option Characteristic Curves}

\begin{figure}[!ht]
\centering
\subfigure[Item 24\label{fig:OCC24 discriminator}]
{\resizebox{0.49\textwidth}{!}{\includegraphics{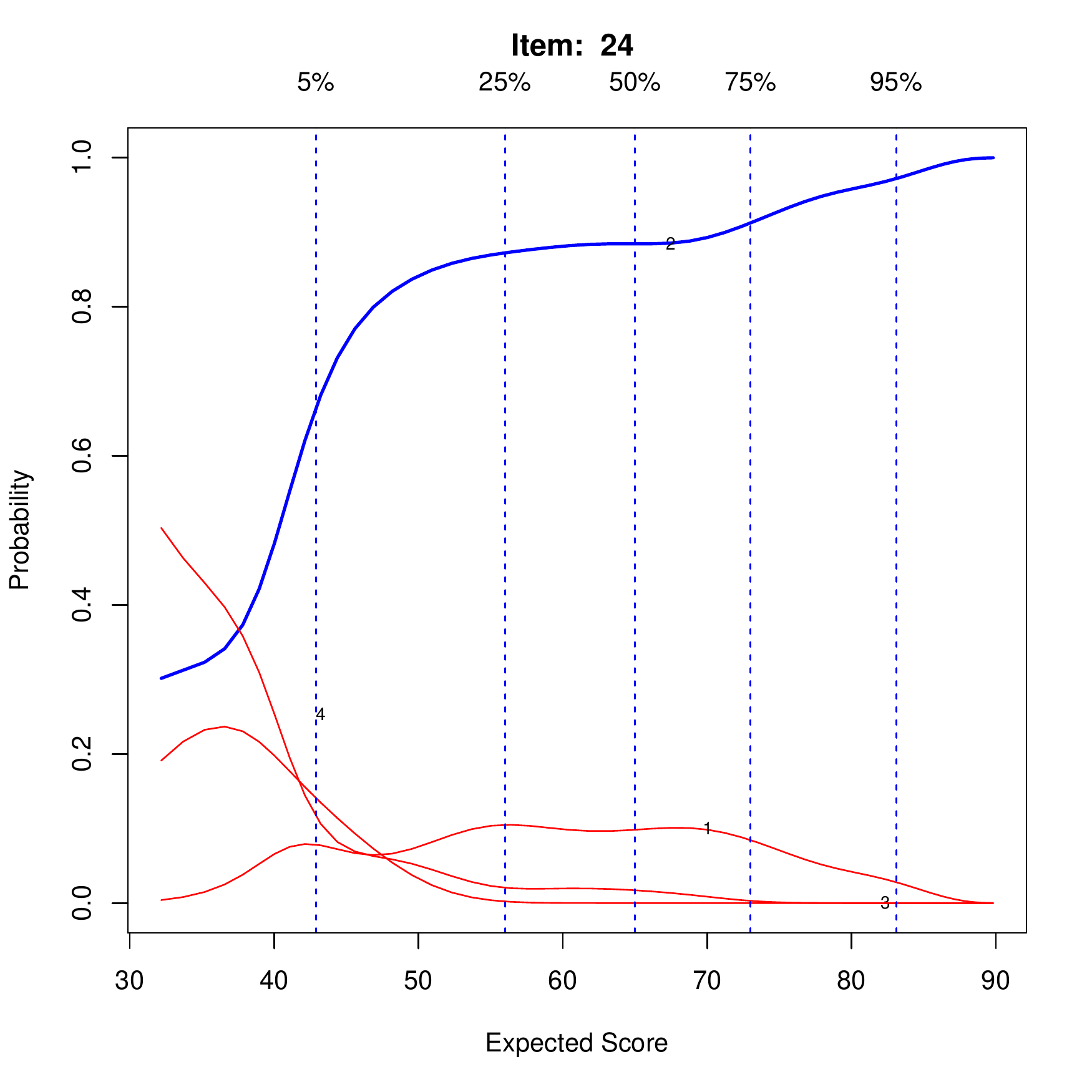}}}
\subfigure[Item 25\label{fig:OCC25 distractor}]
{\resizebox{0.49\textwidth}{!}{\includegraphics{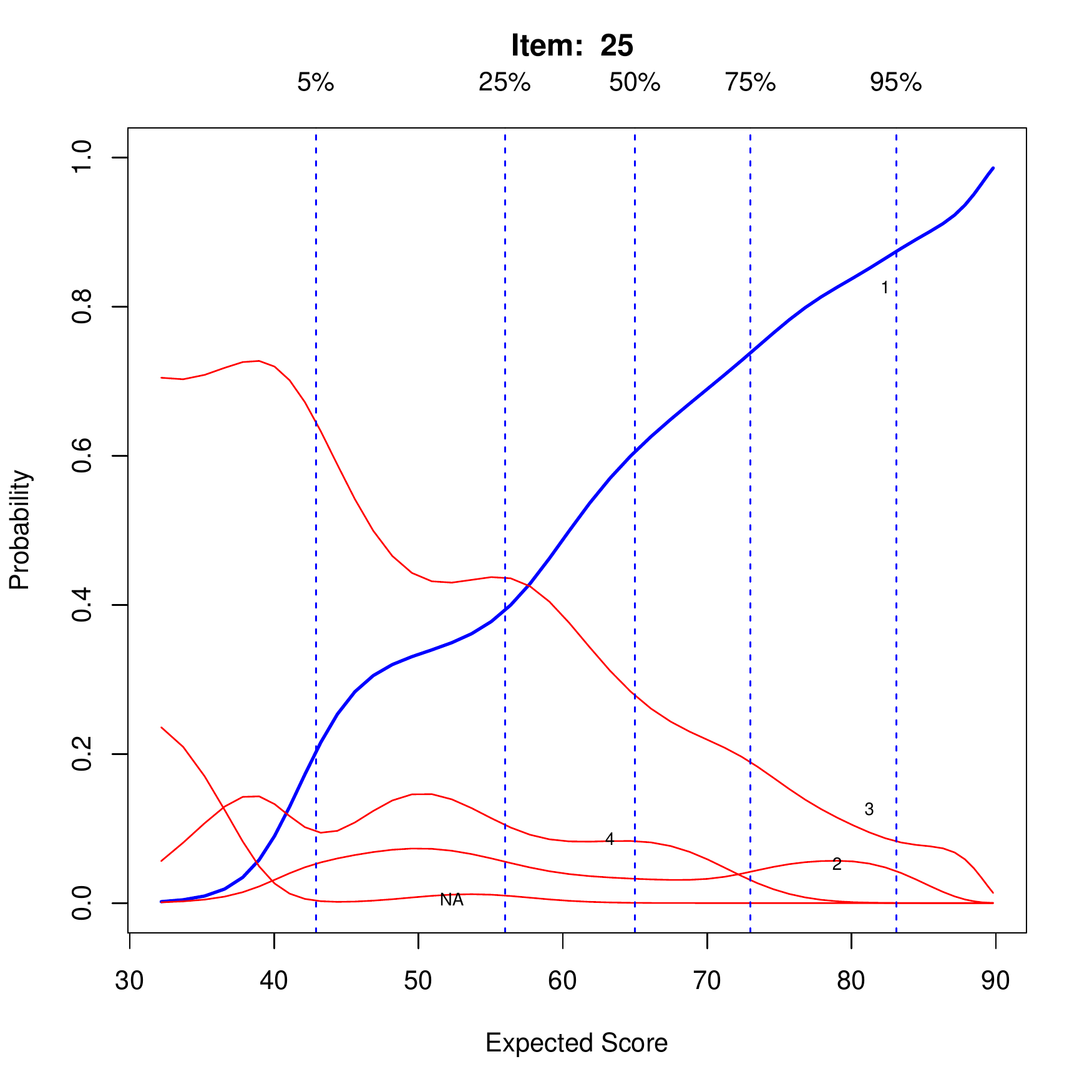}}}
\subfigure[Item 92\label{fig:OCC92 easy}]
{\resizebox{0.49\textwidth}{!}{\includegraphics{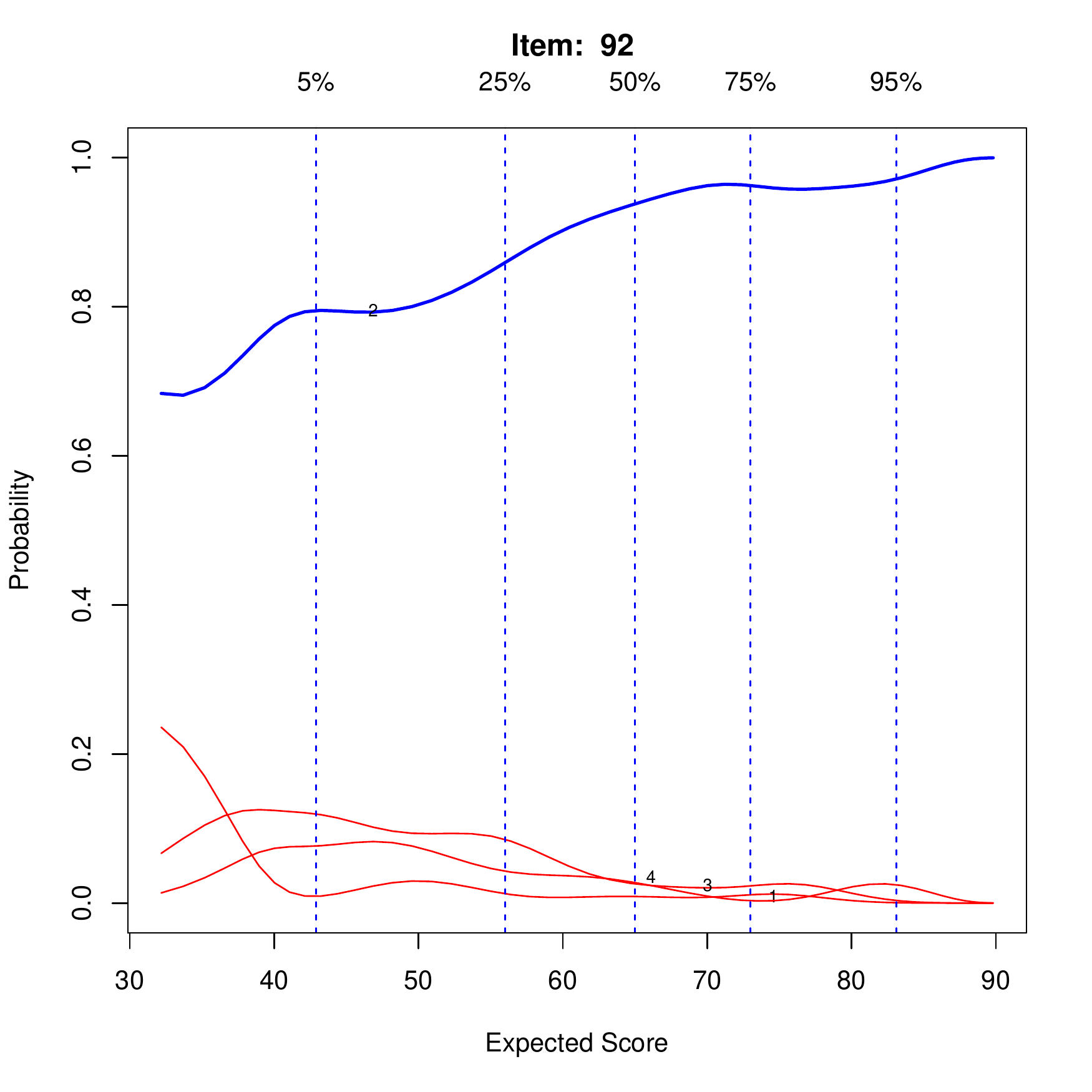}}}
\subfigure[Item 96\label{fig:OCC96 nonmonotone}]
{\resizebox{0.49\textwidth}{!}{\includegraphics{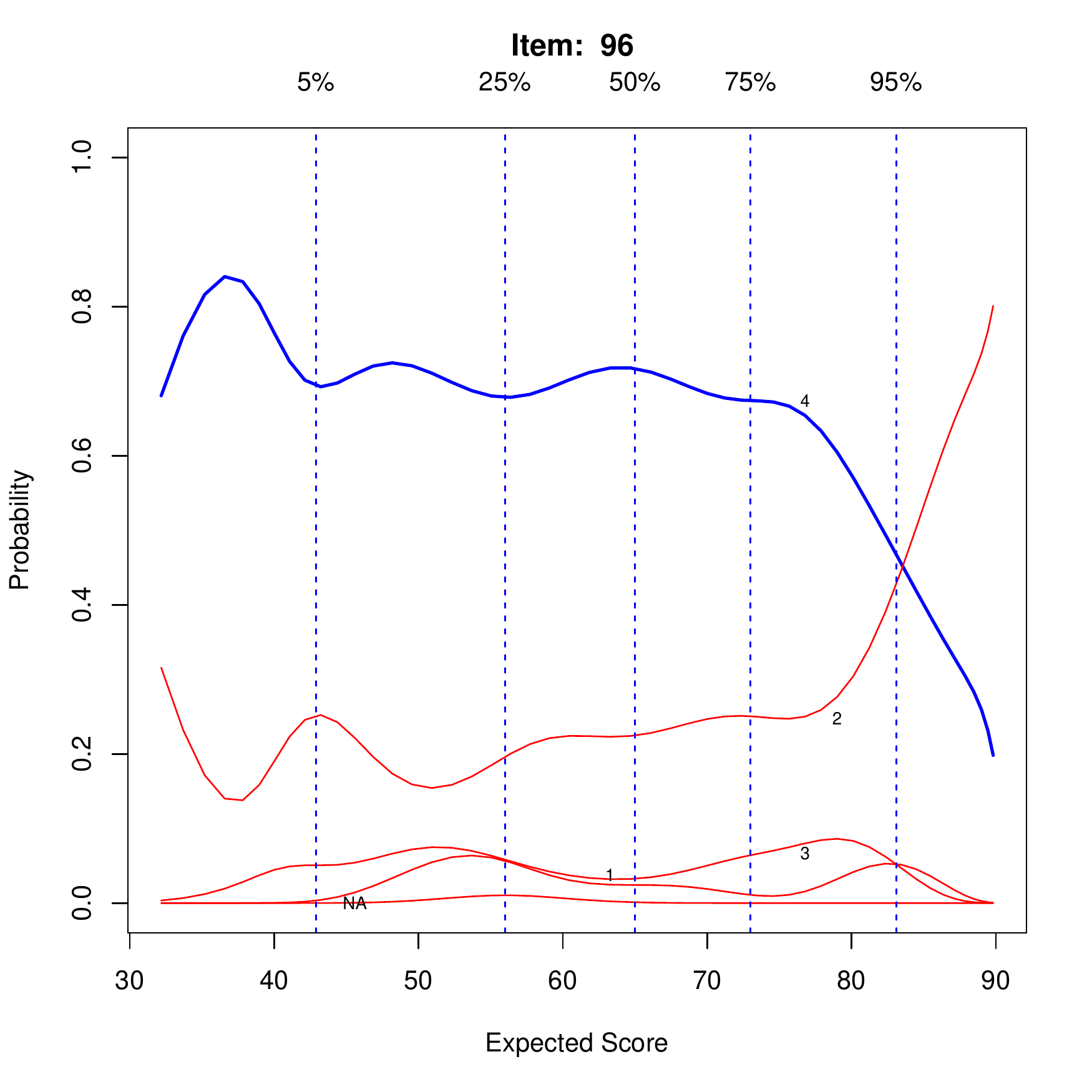}}}
\caption{
OCCs for items 24, 25, 92, and 96 of the introductory psychology exam.
\label{fig:PsychOCCs}
}
\end{figure}

The code 
\begin{CodeInput}
R> plot(Psych1, plottype="OCC", item=c(24,25,92,96))
\end{CodeInput}
produces the OCCs for items 24, 25, 92, and 96 displayed in Figure~\ref{fig:PsychOCCs}.
The correct options are displayed in blue and the incorrect options in red.
The default specification \code{axistype="scores"} uses the expected total score \eqref{eq:expected total score} as display variable on the $x$-axis.
The vertical dashed lines indicate the scores (or quantiles if \code{axistype="distribution"}) below which 5\%, 25\%, 50\%, 75\%, and 95\% of subjects fall.
Since the argument \code{miss} has not been specified, by default (\code{miss="option"}) an additional OCC is plotted for items receiving nonresponses, as we can see from Figure~\ref{fig:OCC25 distractor} and Figure~\ref{fig:OCC96 nonmonotone}.

The OCC plots in Figure~\ref{fig:PsychOCCs} show four very different items.
Globally, apart from item 96 in Figure~\ref{fig:OCC96 nonmonotone}, the other items appear to be monotone enough.
Item 96 is problematic for the Psych 101 instructor, as subjects with lower trait levels are more likely to select the correct option than higher trait level examinees.
In fact, examinees with an expected total score of 90 are the least likely to select the correct option. 
Perhaps the question is misworded or it is measuring a different trait.
On the contrary, items 24, 25, and 92, do a good job in differentiating between subjects with low and high trait levels.
In particular item 24, in Figure~\ref{fig:OCC24 discriminator}, displays a high discriminating power for subjects with expected total scores near 40, and a low discriminating power for subjects with expected total scores greater than 50; above 50, subjects have roughly the same probability of selecting the correct option regardless of their expected total score.
Item 25 in Figure~\ref{fig:OCC25 distractor} is also an effective one, since only the top students are able to recognize option 3 as incorrect; option 3 was selected by about 30.9\% of the test takers, that is the 72.7\% of those who answered incorrectly.
Note also that, for subjects with expected total scores below about 58, option 3 constitutes the most probable choice.
Finally, item 92 in Figure~\ref{fig:OCC92 easy}, aside from being approximately monotone, is also easy, since a subject with expected total score of about 30 already has a 70\% chance of selecting the correct option; only a few examinees are consequently interested to the incorrect options 1, 3, and 4. 

\subsubsection[Expected Item Scores]{Expected item scores}
\label{subsubsec:Item Characteristic Curves}

\begin{figure}[!ht]
\centering
\subfigure[Item 24\label{fig:EIS24 discriminator}]
{\resizebox{0.49\textwidth}{!}{\includegraphics{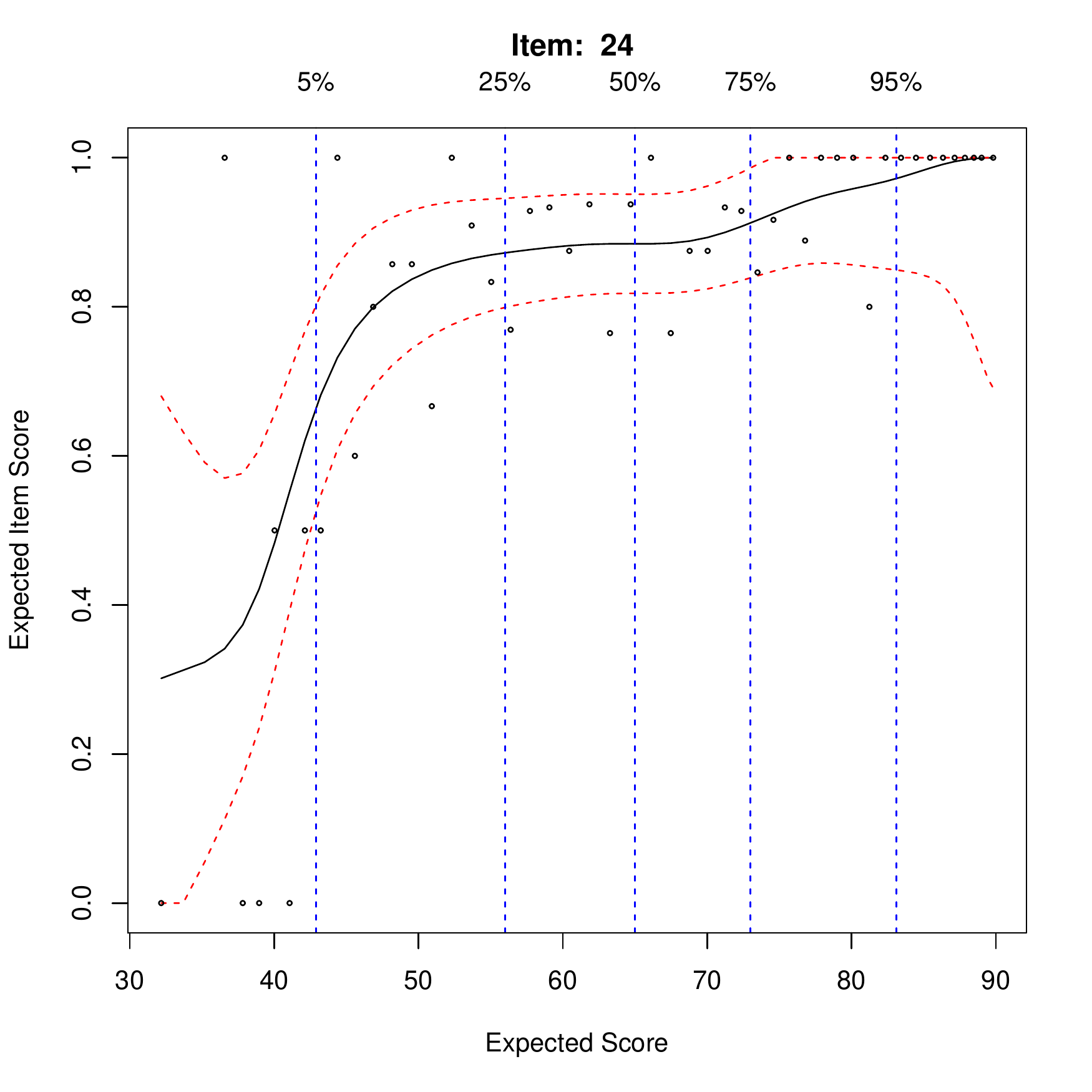}}}
\subfigure[Item 25\label{fig:EIS25 distractor}]
{\resizebox{0.49\textwidth}{!}{\includegraphics{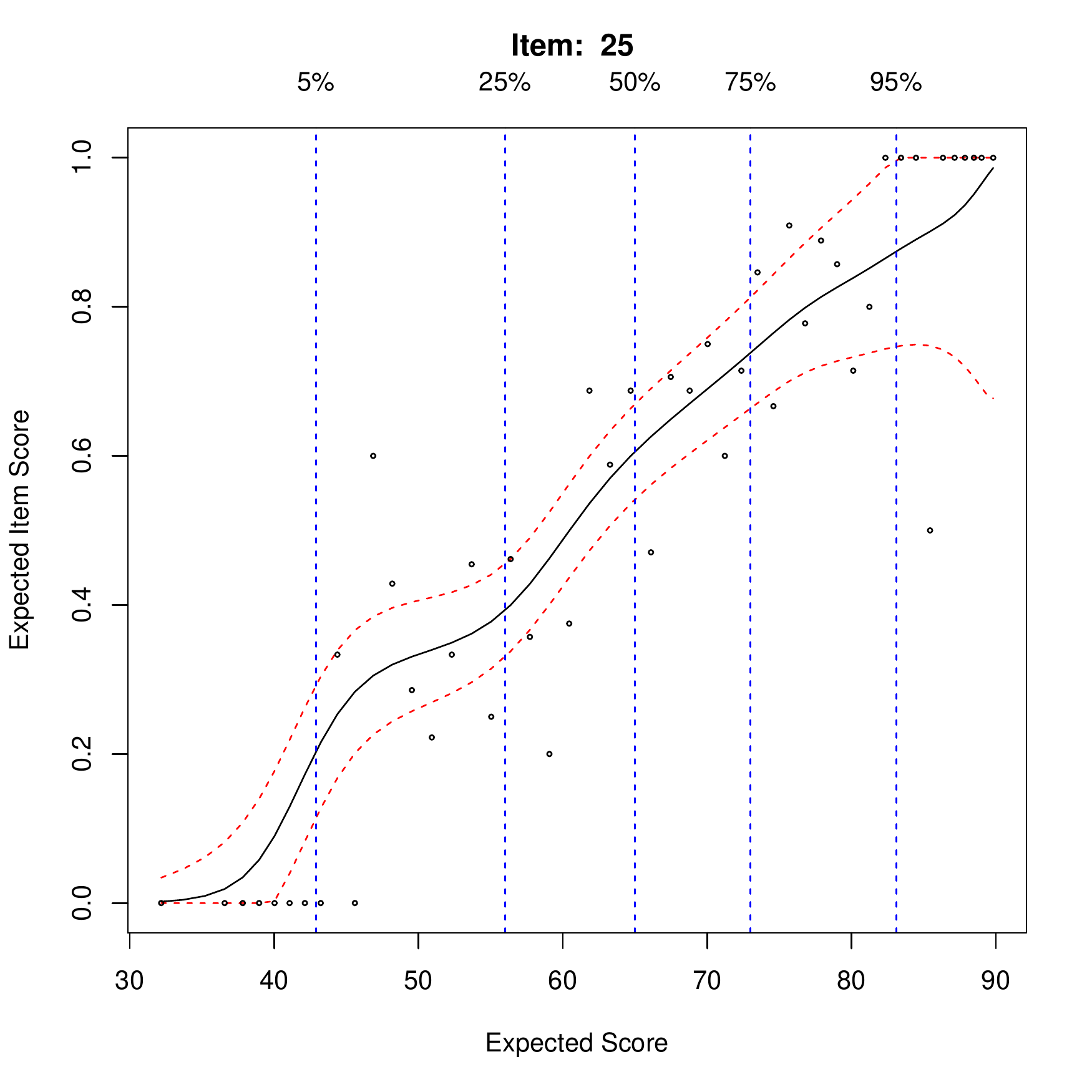}}}
\subfigure[Item 92\label{fig:EIS92 easy}]
{\resizebox{0.49\textwidth}{!}{\includegraphics{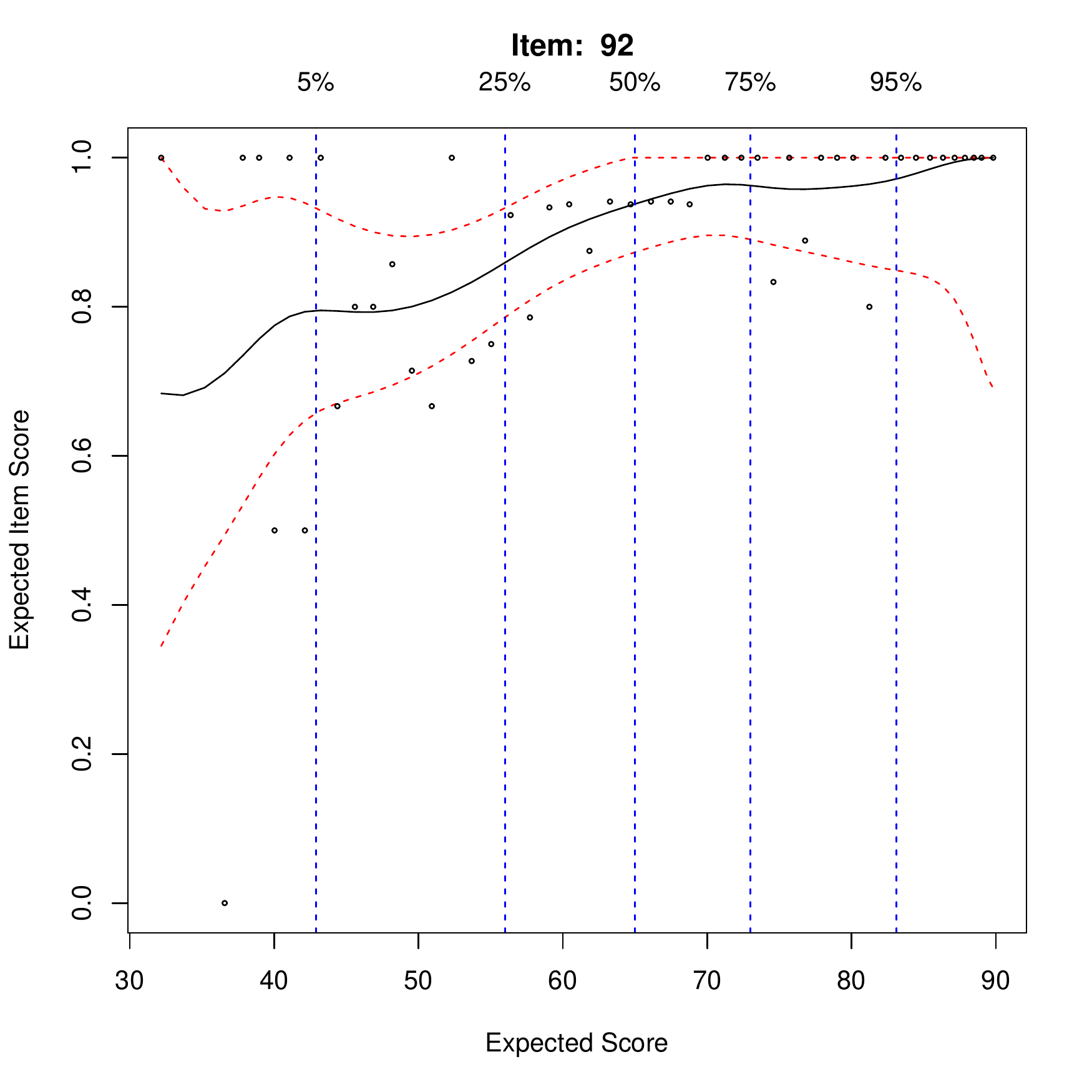}}}
\subfigure[Item 96\label{fig:EIS96 nonmonotone}]
{\resizebox{0.49\textwidth}{!}{\includegraphics{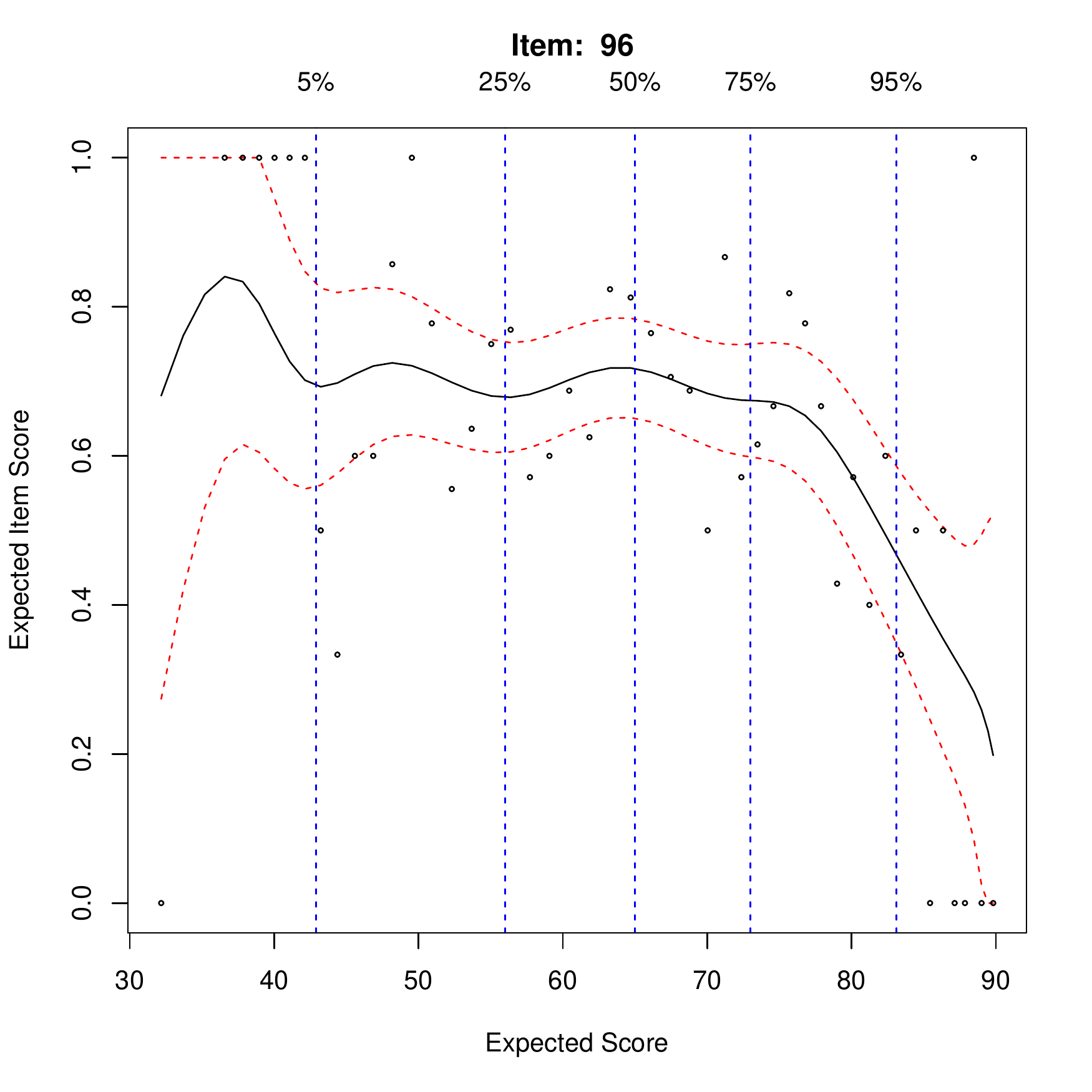}}}
\caption{
EISs, and corresponding 95\% pointwise confidence intervals (dashed red lines), for items 24, 25, 92, and 96 of the introductory psychology exam.
Grouped subject scores are displayed as points. 
\label{fig:PsychEISs}
}
\end{figure}

Through the code 
\begin{CodeInput}
R> plot(Psych1, plottype="EIS", item=c(24,25,92,96))
\end{CodeInput}
we obtain, for the same set of items, the EISs displayed in Figure~\ref{fig:PsychEISs}.
Due to the 0/1 weighting scheme, the EIS is the same as the OCC (shown in blue in Figure~\ref{fig:PsychOCCs}) for the correct option.  
EISs by default show the 95\% approximated pointwise confidence intervals (dashed red lines) illustrated in Section~\ref{subsec:pointwise confidence intervals}.
Via the argument \code{alpha}, these confidence intervals can be removed entirely (\code{alpha=FALSE}) or changed by specifying a different value.
In this example relatively wide confidence intervals, for expected total scores at extremely high or low levels, are obtained. 
This is due to the fact that there are less data for estimating the curve in these regions and thus there is less precision in the estimates.
Finally, the points on the EIS plots show the observed average score for the subjects grouped as in \eqref{eq:practical kernel estimates}.

\subsubsection[Probability simplex plots]{Probability simplex plots}
\label{subsubsec:Probability Simplex Plots}

To complement the OCCs, the package includes triangle and tetrahedron (simplex) plots that, as illustrated in Section~\ref{subsec:Probability Simplex}, synthesize the OCCs.
When these plots are used on items with more than 3 or 4 options (including the missing value category), only the options corresponding to the 3 or 4 highest probabilities will be shown; naturally, these probabilities are normalized in order to allow the simplex representation.
This seldom loses any real information since experience tends to show that in a very wide range of situations people tend to eliminate all but a few options.

\begin{figure}[!ht]
\centering
\subfigure[Item 24\label{fig:triangle 24}]
{\resizebox{0.496\textwidth}{!}{\includegraphics{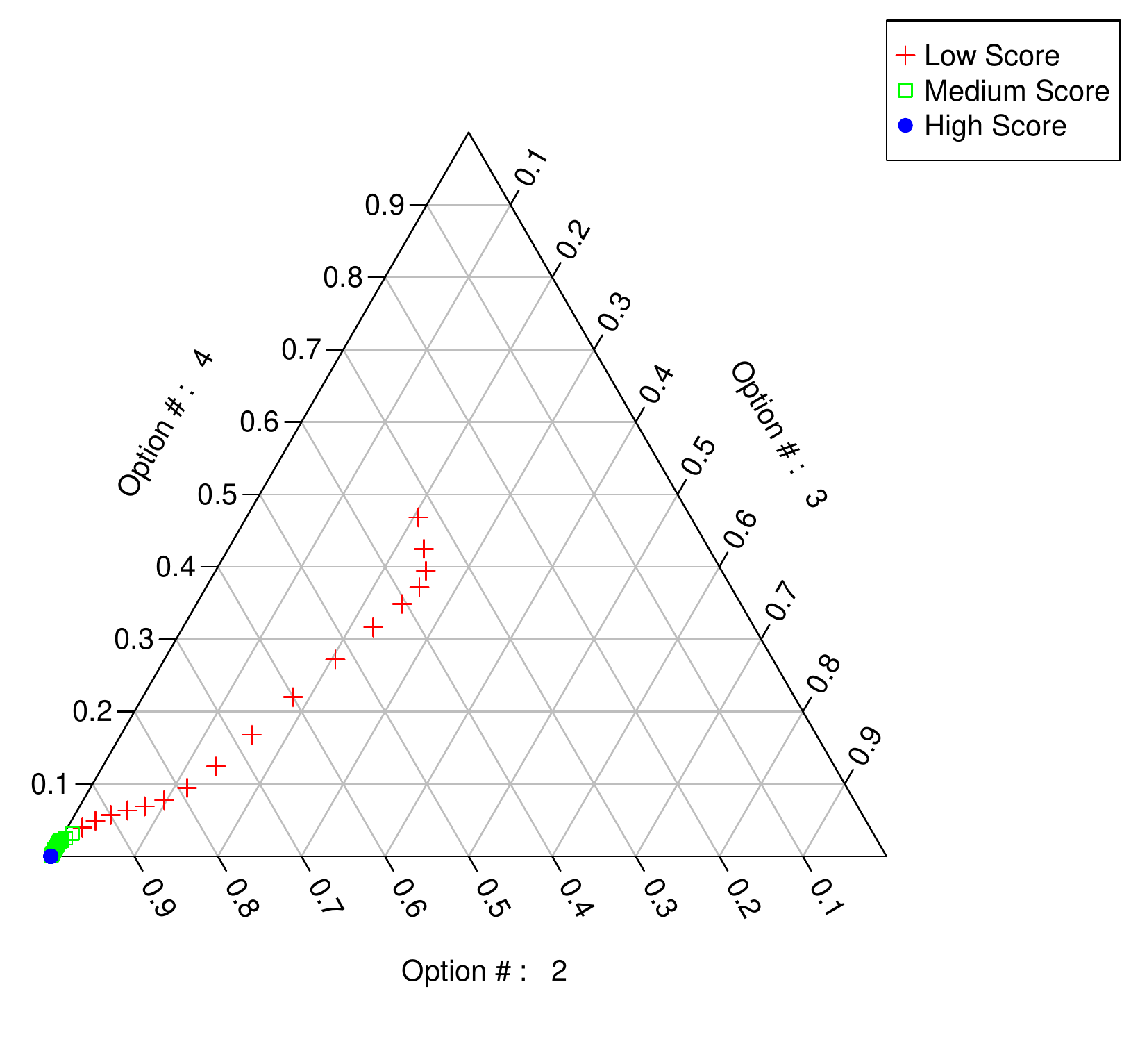}}}
\subfigure[Item 92\label{fig:triangle 92}]
{\resizebox{0.496\textwidth}{!}{\includegraphics{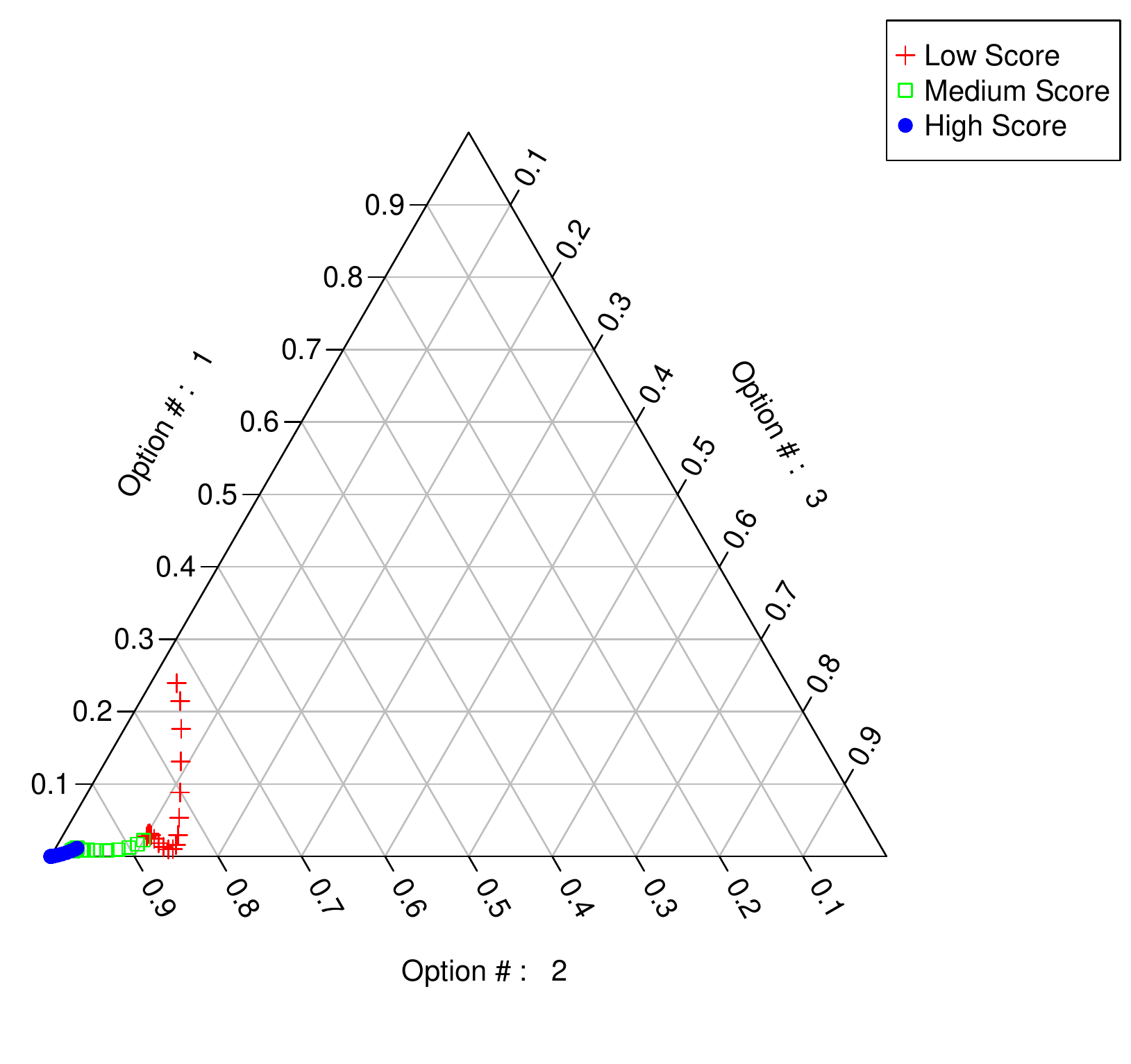}}}
\caption{Probability triangles for two items of the introductory psychology exam. 
\label{fig:triangle psych}
}
\end{figure}

The tetrahedron is the natural choice for the items 24 and 92, characterized by four options and without ``observed'' missing responses; for these items the code 
\begin{CodeInput}
R> plot(Psych1, plottype="tetrahedron", items=c(24,92))
\end{CodeInput}
generates the tetrahedron plots displayed in Figure~\ref{fig:PsychOCCs}.
\begin{figure}[!ht]
\centering
\subfigure[Item 24\label{fig:tetra 24}]
{\resizebox{0.40\textwidth}{!}{\includegraphics{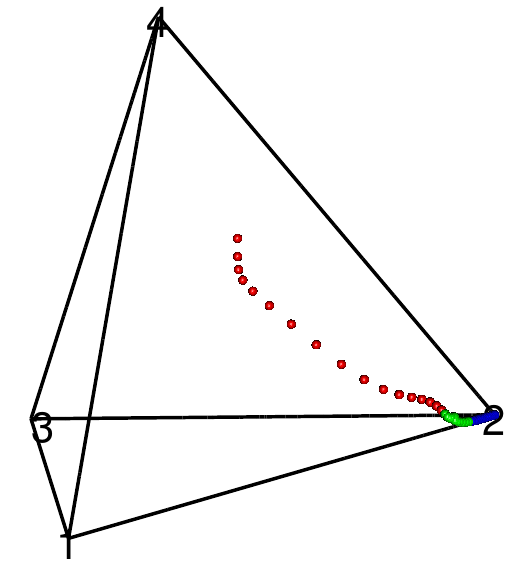}}}
\qquad\qquad
\subfigure[Item 92\label{fig:tetra 92}]
{\resizebox{0.40\textwidth}{!}{\includegraphics{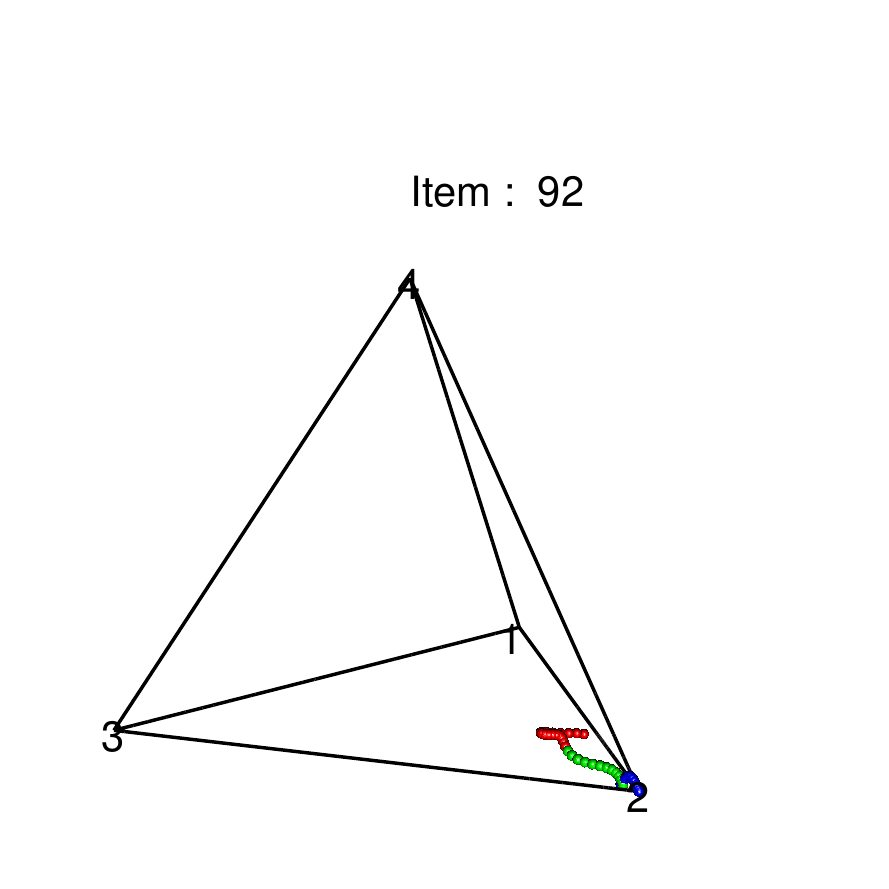}}}
\caption{Probability tetrahedrons for two items of the introductory psychology exam.
Low trait levels are plotted in red, medium in green, and high in blue. 
\label{fig:tetra psych}
}
\end{figure} 
These plots may be manipulated with the mouse or keyboard.
Inside the tetrahedron there is a curve constructed on the $q$ (\code{nevalpoints}) evaluation points. 
In particular, low, medium, and high trait levels are identified by red, green, and blue points, respectively, where the levels are simply the values of \code{evalpoints} broken into three equal groups.
Considering this ordering in the trait level, it is possible to make some considerations.
\begin{itemize}
	\item A basic requirement of a reasonable item, of this format, is that the sequence of points terminates at or near the correct answer.
In these terms, as can be noted in Figure~\ref{fig:tetra 24} and Figure~\ref{fig:tetra 92}, items 24 and 92 satisfy this requirement since the sequence of points moves toward the correct option.
	\item The length of the curve is very important.
	 The individuals with the lowest trait levels should be far from those with the highest.
	 Item 24, in Figure~\ref{fig:tetra 24}, is a fairly good example.
	 By contrast very easy items, such as item 92 in Figure~\ref{fig:tetra 92}, have very short curves concentrated close to the correct answer, with only the worst students showing a slight tendency to choose a wrong answer.
	 \item The spacing of the points is related to the speed at which probabilities of choice change; compare the worst students of Figure~\ref{fig:tetra 24} with those in Figure~\ref{fig:tetra 92} and also the corresponding results in Figure~\ref{fig:OCC24 discriminator} and Figure~\ref{fig:OCC92 easy}, respectively.	 
	 \end{itemize}
For the same items, the code 
\begin{CodeInput}
R> plot(Psych1, plottype="triangle", items=c(24,92))
\end{CodeInput}
produces the triangle plots displayed in Figure~\ref{fig:triangle psych}.
From Figure~\ref{fig:triangle 24} we can see that in the set of the three most chosen options, the second one has a much higher probability of being selected while the other two share almost the same 
probability, and so the sequence of points approximately lies on the bisector of the angle associated to the second option.

\subsubsection[Principal component analysis]{Principal component analysis}
\label{subsubsec:Psych - Principal Component Analysis}

\begin{figure}[!ht]
\centering
{\includegraphics[totalheight=0.6\textheight]{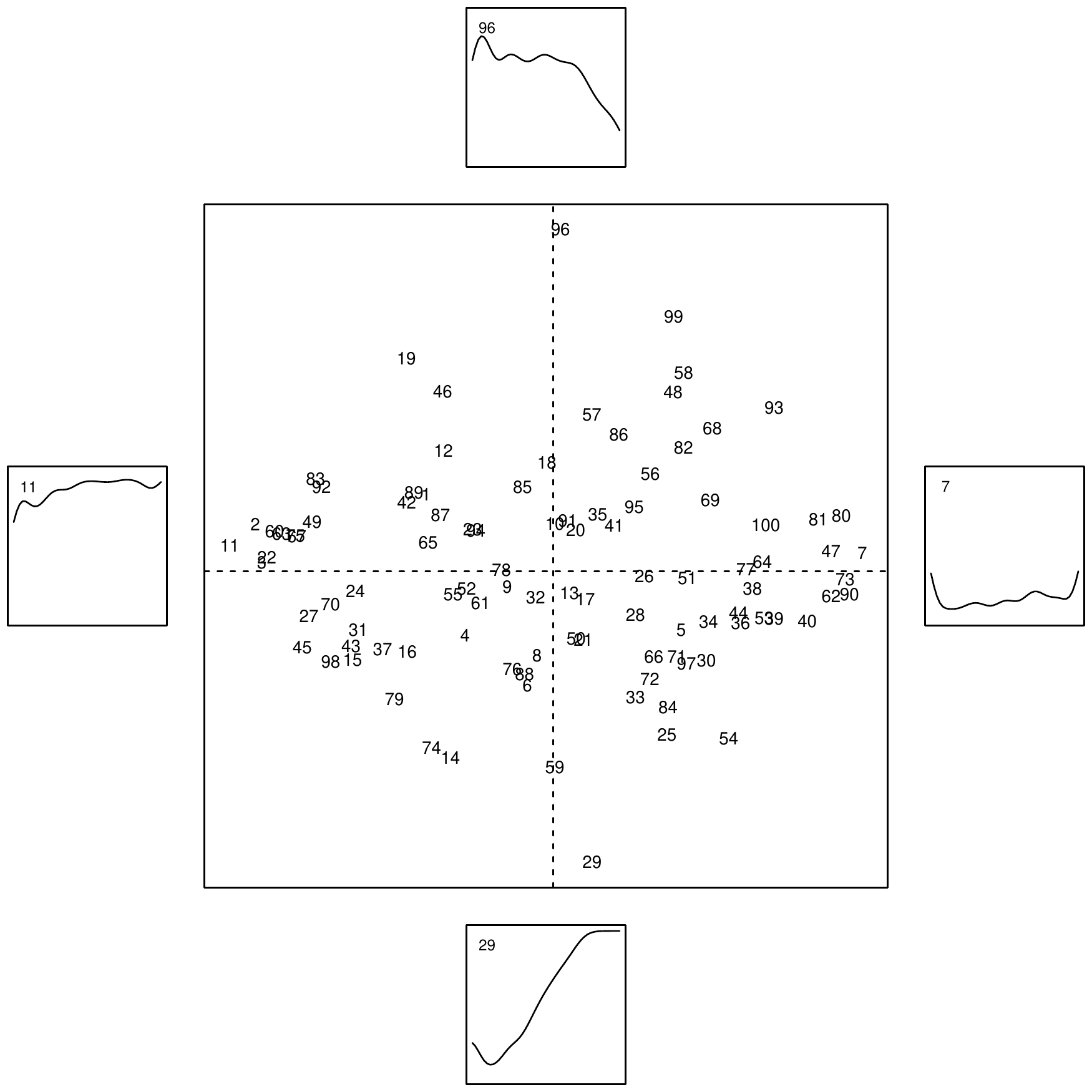}}
\caption{
First two principal components for the introductory psychology exam.
In the interior plot, numbers are the identifiers of the items.
The vertical component represents discrimination, while the horizontal one difficulty.
The small plots show the EISs for the most extreme items for each principal component.
\label{fig:PCA psych} 
}
\end{figure}

By performing a principal component analysis (PCA) of the EISs at each evaluation point, the \pkg{KernsmoothIRT} package provides a way to simultaneously compare items and to show the relationships among them.
Since EISs may be defined on different ranges $\left[x_{j\min},x_{j\max}\right]$, the transformation $\left(\widehat{e}_j\left(\vartheta\right)-x_{j\min}\right)/\left(x_{j\max}-x_{j\min}\right)$, $j=1,\ldots,k$, is preliminary applied. 
Furthermore, as stated in Section~\ref{sec:Estimating abilities}, in this paradigm only rank order considerations make sense, so the zero-centered ranks of $\widehat{e}_1\left(\vartheta_s\right),\ldots,\widehat{e}_k\left(\vartheta_s\right)$, for each $s=1,\ldots,q$, are computed and the PCA is carried out on the resulting $\left(q \times k\right)$-matrix.
In particular, the code
\begin{CodeInput}
R> plot(Psych1, plottype="PCA")
\end{CodeInput}
produces the graphical representation in Figure~\ref{fig:PCA psych}.
A first glance to this plot shows that:
\begin{itemize}
	\item the first principal component, on the horizontal axis, represents item difficulty, since the most difficult items are placed on the right and the easiest ones on the left.
	The small plots on the left and on the right show the EISs for the two extreme items with respect to this component and help the user in identifying the axis-direction with respect to difficulty (from low to high or from high to low).
	Here, $I_7$ shows high difficulty, as test takers of all ability levels receive a low score, while $I_{11}$ is extremely easy.
	\item the second principal component, on the vertical axis, corresponds to item discrimination, since low items tend to have an high positive slope while high items tend to have an high negative slope.
	Also in this case, the small plots on the bottom and on the top show the EISs for the two extreme items with respect to this component and help the user in identifying the axis-direction with respect to discrimination (from low to high or \textit{viceversa}).
	Here, while both $I_{96}$ and $I_{29}$ possess a very strong discrimination, $I_{96}$ is clearly ill-posed, since it discriminates negatively.
	\end{itemize}
Concluding, the principal components plot tends to be a useful overall summary of the composition of the test.
Figure~\ref{fig:PCA psych} is fairly typical of most academic tests and it is also usual to have only two dominant principal components reflecting item difficulty and discrimination.

\subsubsection[Relative Credibility Curves]{Relative credibility curves}
\label{subsubsec:Psych - Relative Credibility Plot}

\begin{figure}[!ht]
\centering
\subfigure[Subject 13\label{fig:PsychRCC13}]
{\resizebox{0.49\textwidth}{!}{\includegraphics{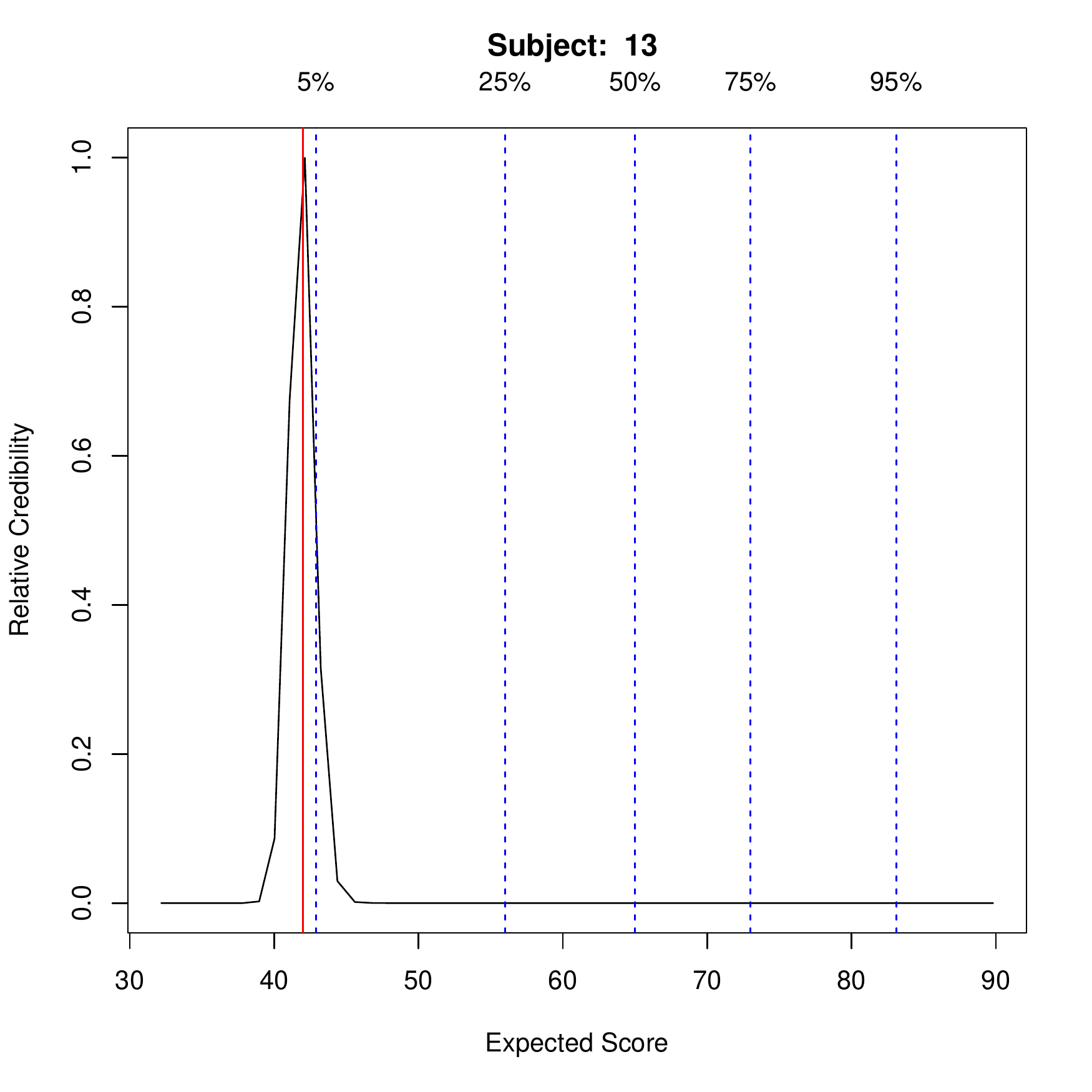}}}
\subfigure[Subject 92\label{fig:PsychRCC92}]
{\resizebox{0.49\textwidth}{!}{\includegraphics{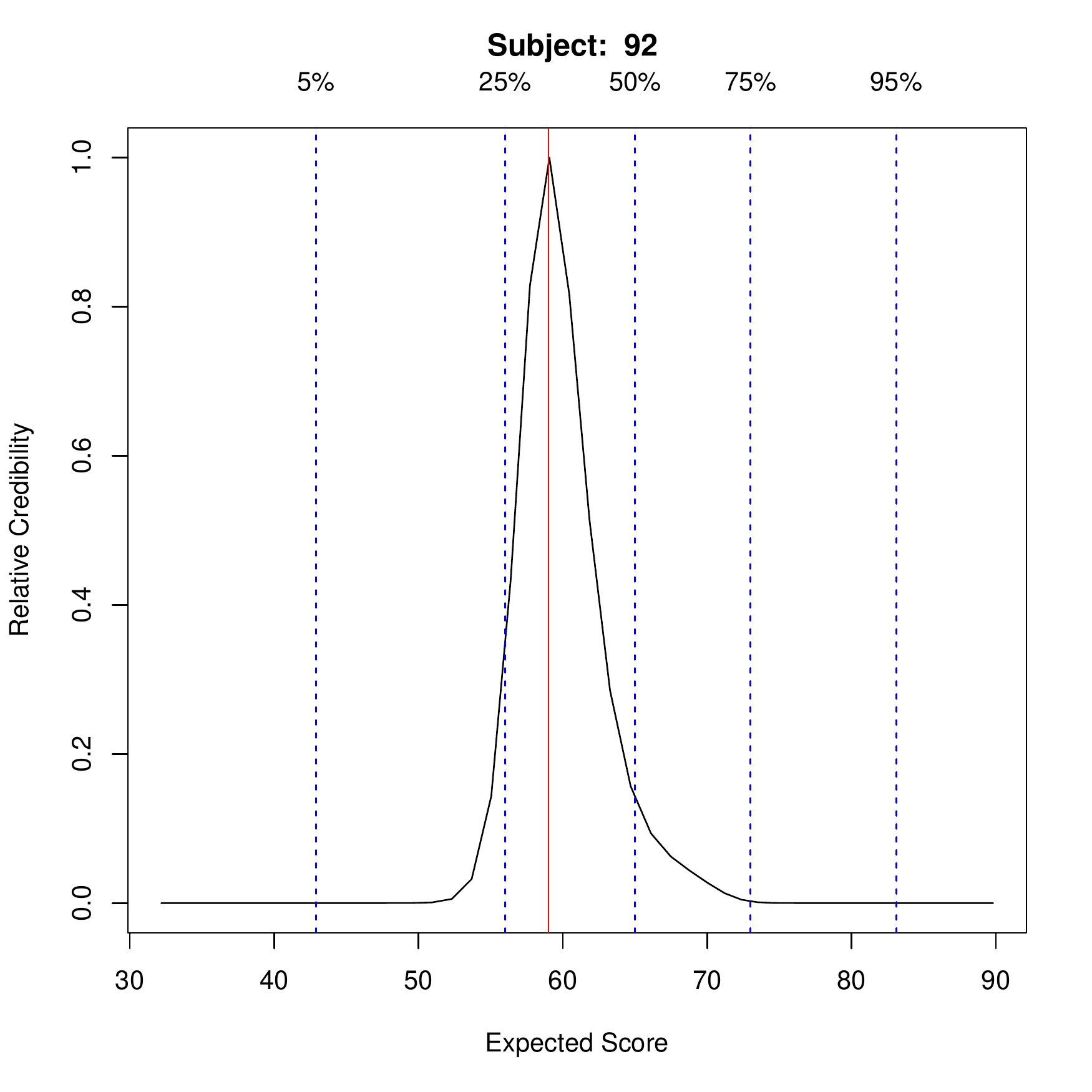}}}
\subfigure[Subject 111\label{fig:PsychRCC111}]
{\resizebox{0.49\textwidth}{!}{\includegraphics{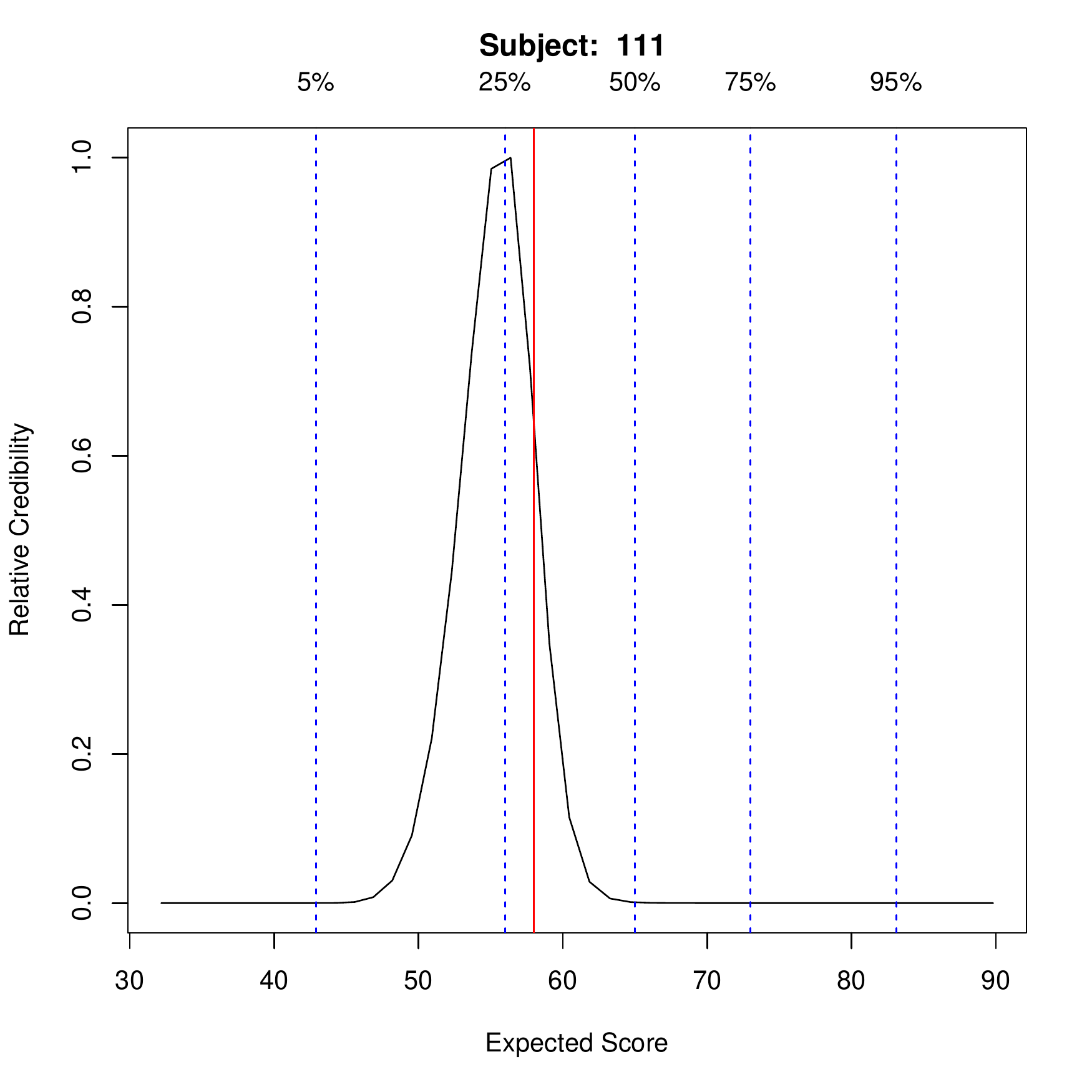}}}
\subfigure[Subject 33\label{fig:PsychRCC33}]
{\resizebox{0.49\textwidth}{!}{\includegraphics{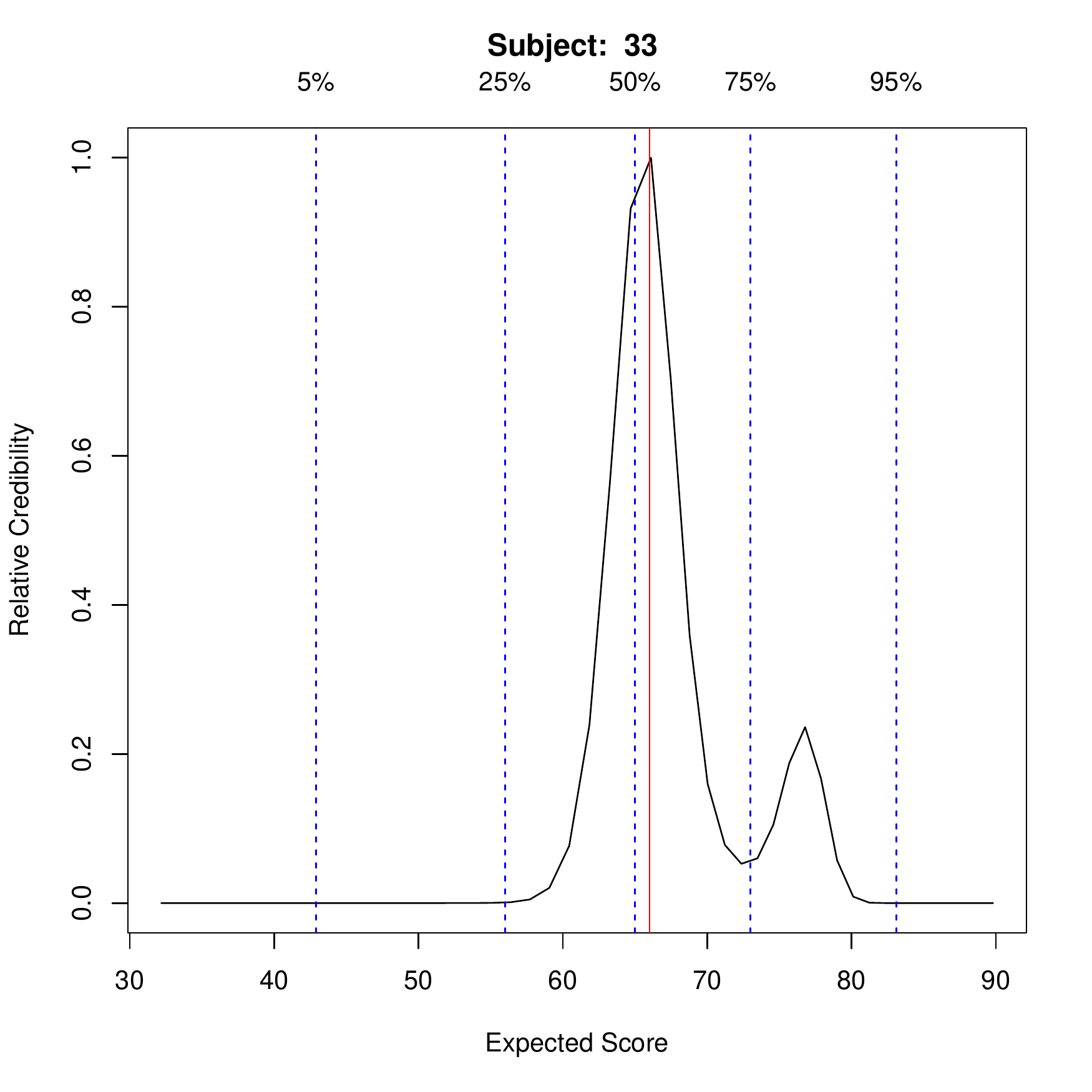}}}
\caption{RCCs for some subjects. 
The vertical red line shows the actual score the subject received. 
\label{fig:Psych RCC}
}
\end{figure}

The RCCs shown in Figure~\ref{fig:Psych RCC} are obtained by the command
\begin{CodeInput}
R> plot(Psych1, plottype="RCC", subjects=c(13,92,111,33))
\end{CodeInput}
In each plot, the red line shows the subject's actual score $t$.

For both the subjects considered in Figure~\ref{fig:PsychRCC13} and Figure~\ref{fig:PsychRCC92}, there is a substantial agreement between the maximum of the RCC, $\widehat{e}\left(\widehat{\vartheta}^{\text{ML}}\right)$, and $t$.
Nevertheless, there is a difference in terms of the precision of the ML-estimates; for $S_{13}$ the RCC is indeed more spiky, denoting a higher precision.
In Figure~\ref{fig:PsychRCC111} there is a substantial difference between $\widehat{e}\left(\widehat{\vartheta}^{\text{ML}}_{111}\right)$ and $t_{111}$.
This indicates that the correct and incorrect answers of this subject are more consistent with a lower score than they are with the actual score received.
Finally, in Figure~\ref{fig:PsychRCC33}, although there is a substantial agreement between $\widehat{e}\left(\widehat{\vartheta}^{\text{ML}}_{33}\right)$ and $t_{33}$, a small but prominent bump is present in the right part of the plot.
Although $S_{33}$ is well represented by his total score, he passed some, albeit few, difficult items and this may lead to think that he is more able than $t_{33}$ suggests.

The commands
\begin{CodeChunk}
\begin{CodeInput}
R> subjscore(Psych1)
\end{CodeInput}
\begin{CodeOutput}
[1] 74 56 89 70 56 57 ...
\end{CodeOutput}
\begin{CodeInput}
R> subjscoreML(Psych1)
\end{CodeInput}
\begin{CodeOutput}
[1] 72.36589 59.06626 88.47615 67.47167 57.71787 55.03844 ...
\end{CodeOutput}
\end{CodeChunk}
allow us to evaluate the differences between the values of $t_i$ and $\widehat{e}\left(\widehat{\vartheta}^{\text{ML}}_i\right)$, $i=1,\ldots,n$.

\subsubsection[Test summary plots]{Test summary plots}
\label{subsubsec: Test Summary Plots}

\begin{figure}[!ht]
\centering
\subfigure[Expected Test Score\label{fig:Test Characteristic Function}]
{\resizebox{0.328\textwidth}{!}{\includegraphics{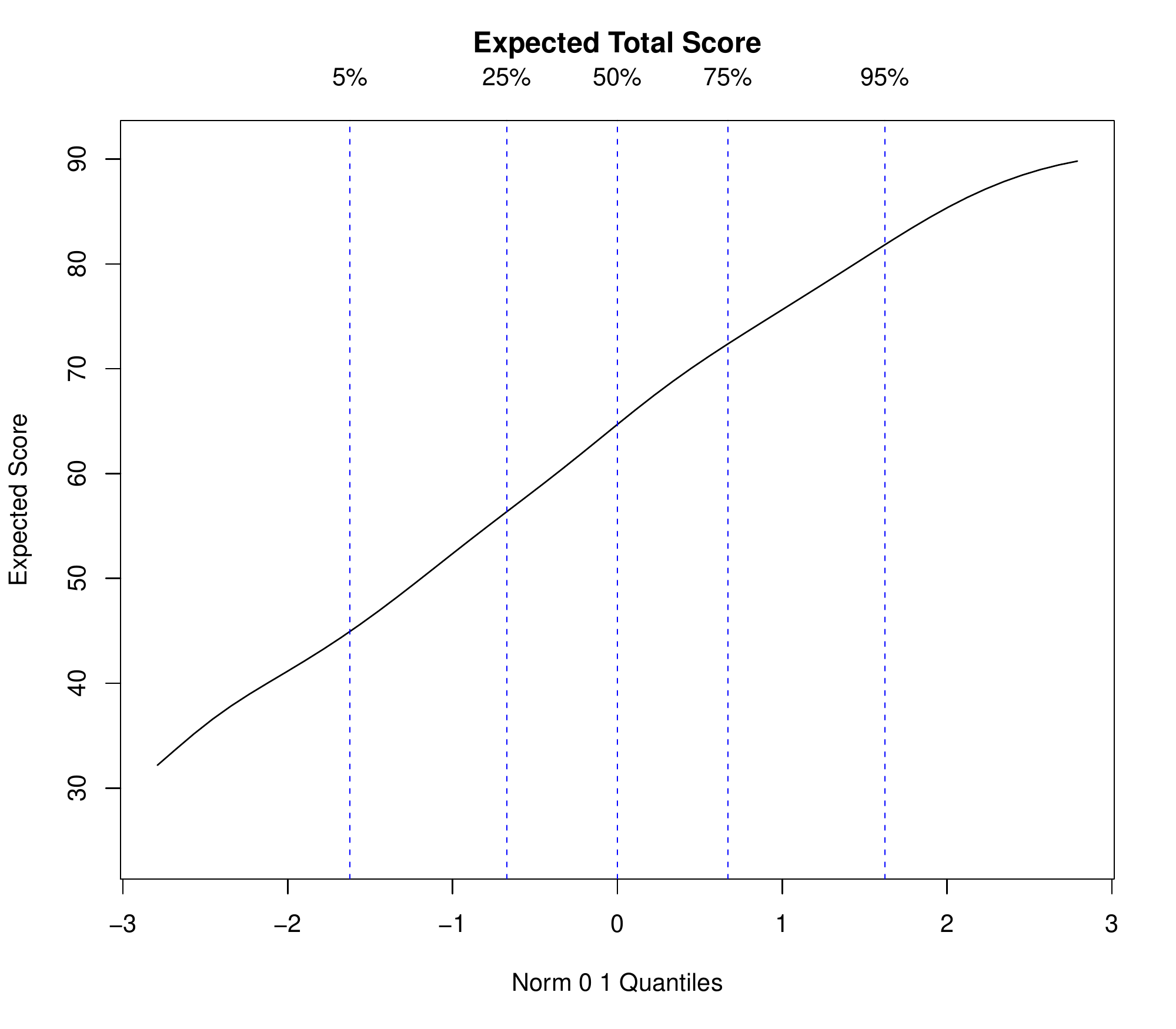}}}
\subfigure[Standard Deviation\label{fig:Standard Deviation}]
{\resizebox{0.328\textwidth}{!}{\includegraphics{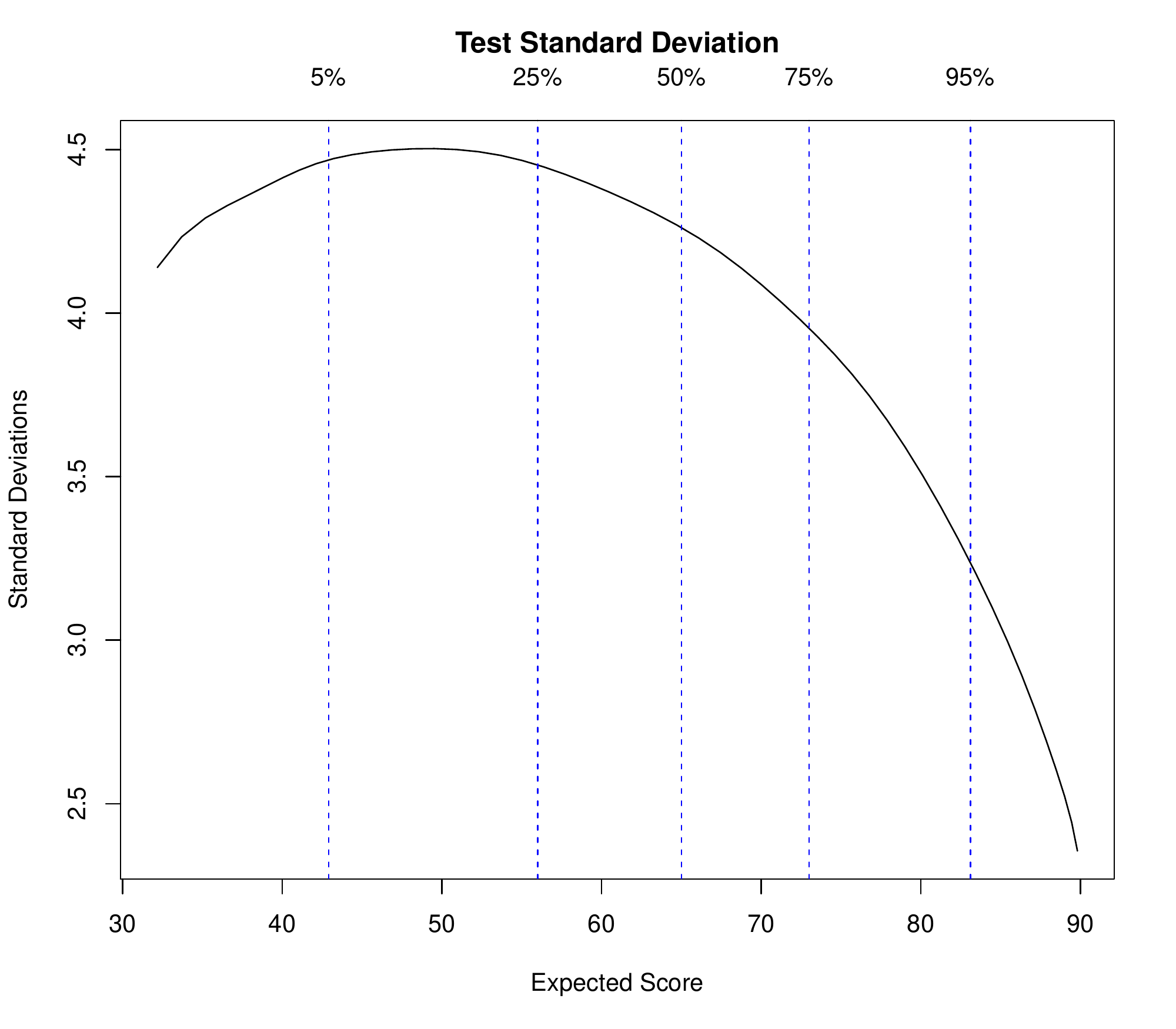}}}
\subfigure[Density\label{fig:Density}]
{\resizebox{0.328\textwidth}{!}{\includegraphics{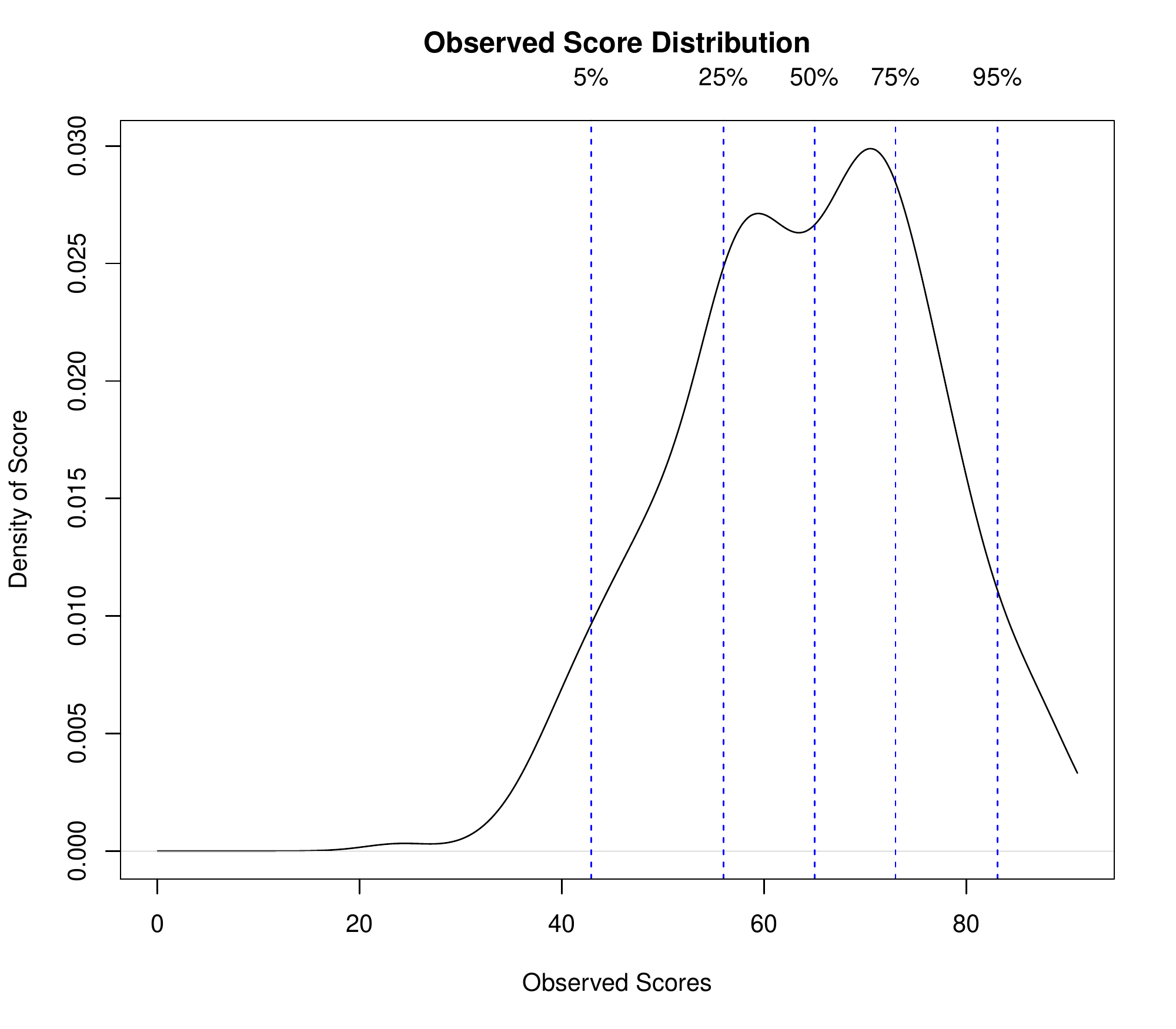}}}
\caption{Test summary plots for the introductory psychology exam.
\label{fig:TestSum} 
}
\end{figure}

The \pkg{KernSmoothIRT} package also contains many analytical tools to assess the test overall. 
Figure~\ref{fig:TestSum} shows a few of these, obtained via the code
\begin{CodeInput}
R> plot(Psych1, plottype="expected")
R> plot(Psych1, plottype="sd")
R> plot(Psych1, plottype="density")
\end{CodeInput}
Figure~\ref{fig:Test Characteristic Function} shows the ETS as a function of the quantiles of the standard normal distribution $\Phi$; it is nearly linear for the Psych 101 dataset.
Note that, in the nonparametric context, the ETS may be non-monotone due to either ill-posed items or random variations.
In the latter case, a slight increase of the bandwidth may be advisable.  

The total score, for subjects having a particular value $\vartheta$, is a random variable, in part because different examinees, or even the same examinee on different occasions, cannot be expected to make exactly the same choices.
The standard deviation of these values, graphically represented in Figure~\ref{fig:Standard Deviation}, is therefore also a function of $\vartheta$.
Figure~\ref{fig:Standard Deviation} indicates that the standard deviation reaches the maximum for examinees at around a total score of $50$, where it is about 4.5 items out of 100.
This translates into 95\% confidence limits of about 41 and 59 for a subject getting 50 items correct.

Figure~\ref{fig:Density} shows a kernel density estimate of the distribution of the total score. 
Although such distribution is commonly assumed to be ``bell-shaped'', from this plot we can note as this assumption is strong for these data.
In particular, a negative skewness can be noted which is a consequence of the test having relatively more easy items than hard ones.
Moreover, bimodality is evident.

\subsection[Voluntary HIV-1 counseling and testing efficacy study group]{Voluntary HIV-1 counseling and testing efficacy study group}
\label{subsec:Voluntary HIV-1}

It is often useful to explore if, for a specific item on a test, its expected score differs when estimated on two or more different groups of subjects, commonly formed by gender or ethnicity.
This is called Differential Item Functioning (DIF) analysis in the psychometric literature.
In particular, DIF occurs when subjects with the same ability but belonging to different groups have a different probability of choosing a certain option.
DIF can properly be called \textit{item bias} because the curves of an item should depend only on $\vartheta$, and not directly on other person factors. 
\citet{Zumb:Thre:2007} offers a recent review of various DIF detection methods and strategies.

The \pkg{KernSmoothIRT} package allows for a nonparametric graphical analysis of DIF, based on kernel smoothing methods.
To illustrate this analysis, we use data coming from the Voluntary HIV Counseling and Testing Efficacy Study, conducted in 1995-1997 by the Center for AIDS Prevention Studies at University of California, San Francisco \citep[see][for details]{Thev:Eff:2000,Thev:Thev:2000}. 
This study was concerned with the effectiveness of HIV counseling and testing in reducing risk behavior for the sexual transmission of HIV.
To perform this study, $n=4292$ persons were enrolled.
The whole dataset -- downloadable from \url{http://caps.ucsf.edu/research/datasets/}, which also contains other useful survey details -- reported 1571 variables for each participant.
As part of this study, respondents were surveyed about their attitude toward condom use via a bank of $k=15$ items.
Respondents were asked how much they agreed with each of the statements on a 4-point response scale, with 1=``strongly disagree'', 2=``disagree more than I agree'', 3=``agree more than I disagree'', 4=``strongly agree'').
Since 10 individuals omitted all the 15 questions, they have been preliminary removed from the used data.
Moreover, given the (``negative'') wording of the items $I_2$, $I_3$, $I_5$, $I_7$, $I_8$, $I_{11}$, and $I_{14}$, a respondent who strongly agreed with such statements was indicating a less favorable attitude toward condom use.
In order to uniform the data, the score for these seven items was preliminary reversed.  
The dataset so modified can be directly loaded from the \pkg{KernSmoothIRT} package by the code
\begin{CodeChunk}
\begin{CodeInput}
R> data("HIV")
R> HIV
\end{CodeInput}
\begin{CodeOutput}
     SITE GENDER AGE  1  2  3  4  5  6  7  8  9  10  11  12  13  14  15
   1  Ken      F  17  4  1  1  4  1  2  4  4  4   4   3   4   1   2   4
   2  Ken      F  17  4  2  4  4  2  3  1  4  3   3   2   3   4   1   4
   3  Ken      F  18  4  4  4  4  4  1  4  4  4   1  NA   4   1  NA   4
   .    .      .   .  .  .  .  .  .  .  .  .  .   .   .   .   .   .   .
   .    .      .   .  .  .  .  .  .  .  .  .  .   .   .   .   .   .   .
   .    .      .   .  .  .  .  .  .  .  .  .  .   .   .   .   .   .   .
4281  Tri      M  79  4  4  1  4  1 NA  4 NA  4  NA  NA  NA  NA   1   4
4282  Tri      M  80  4 NA  4  4  1  4 NA NA NA   1  NA   4   1   4  NA     
\end{CodeOutput}
\begin{CodeInput}
R> attach(HIV)
\end{CodeInput}
\end{CodeChunk}
As it can be easily seen, the above data frame contains the following person factors:
\begin{center}
\begin{tabular}{rcl}
\code{SITE} &=& ``site of the study'' (\code{Ken}=Kenya, \code{Tan}=Tanzania, \code{Tri}=Trinidad) \\
\code{GENDER} &=& ``subject's gender'' (\code{M}=male, \code{F}=female)\\
\code{AGE} &=& ``subject's age'' (age at last birthday)\\
\end{tabular}
\end{center}
Each of these factors can potentially be used for a DIF analysis.
These data have been also analyzed, through some well-known parametric models, by \citet{Bert:Musc:Punz:Item:2010} which also perform a DIF analysis.
Part of this sub-questionnaire has been also considered by \citet{DeAy:Thee:2003,DeAy:Thet:2009} with a Rasch Analysis. 

\begin{figure}[!ht]
\centering
\subfigure[OCCs \label{fig:OCC 9}]
{\resizebox{0.49\textwidth}{!}{\includegraphics{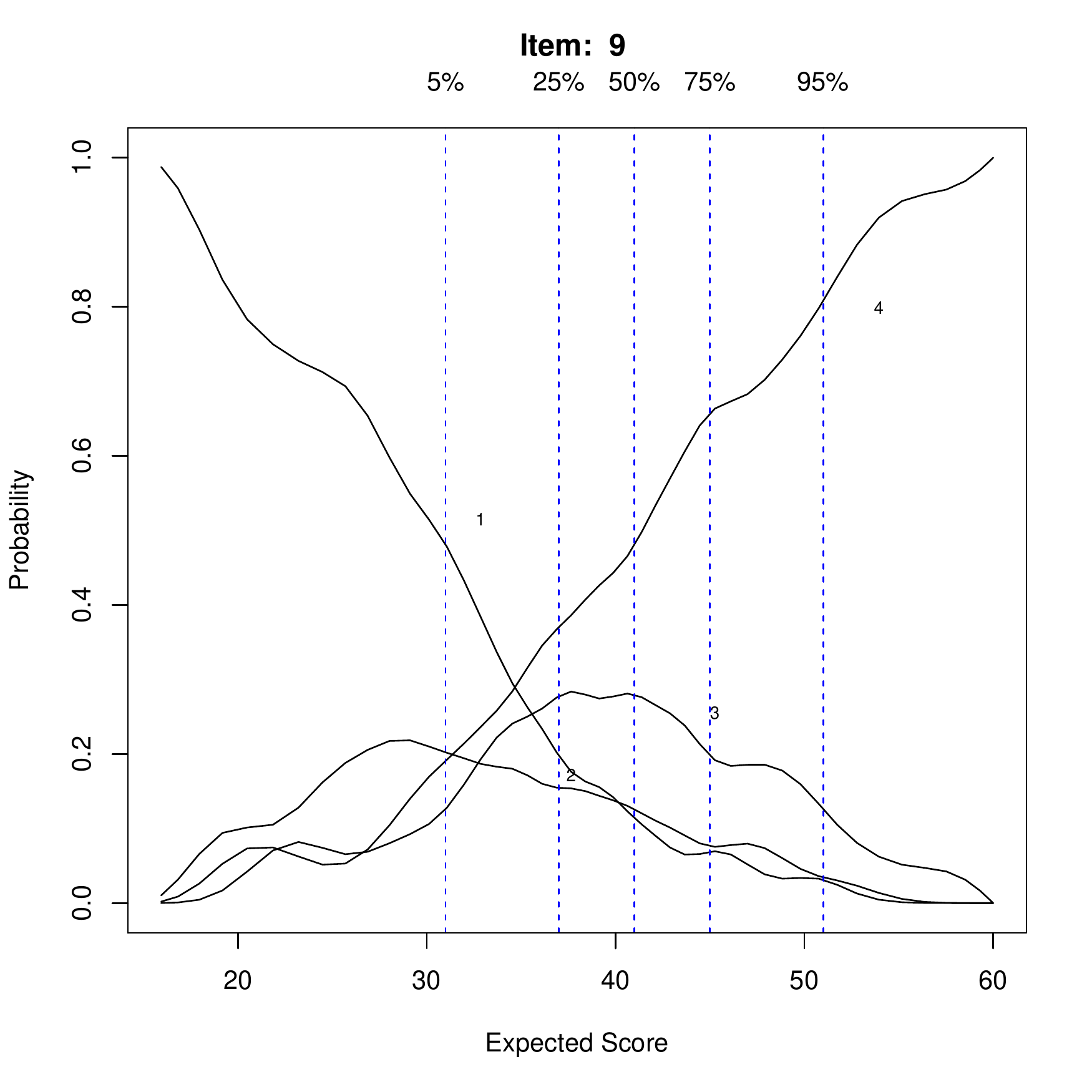}}}
\subfigure[EIS \label{fig:EIS 9}]
{\resizebox{0.49\textwidth}{!}{\includegraphics{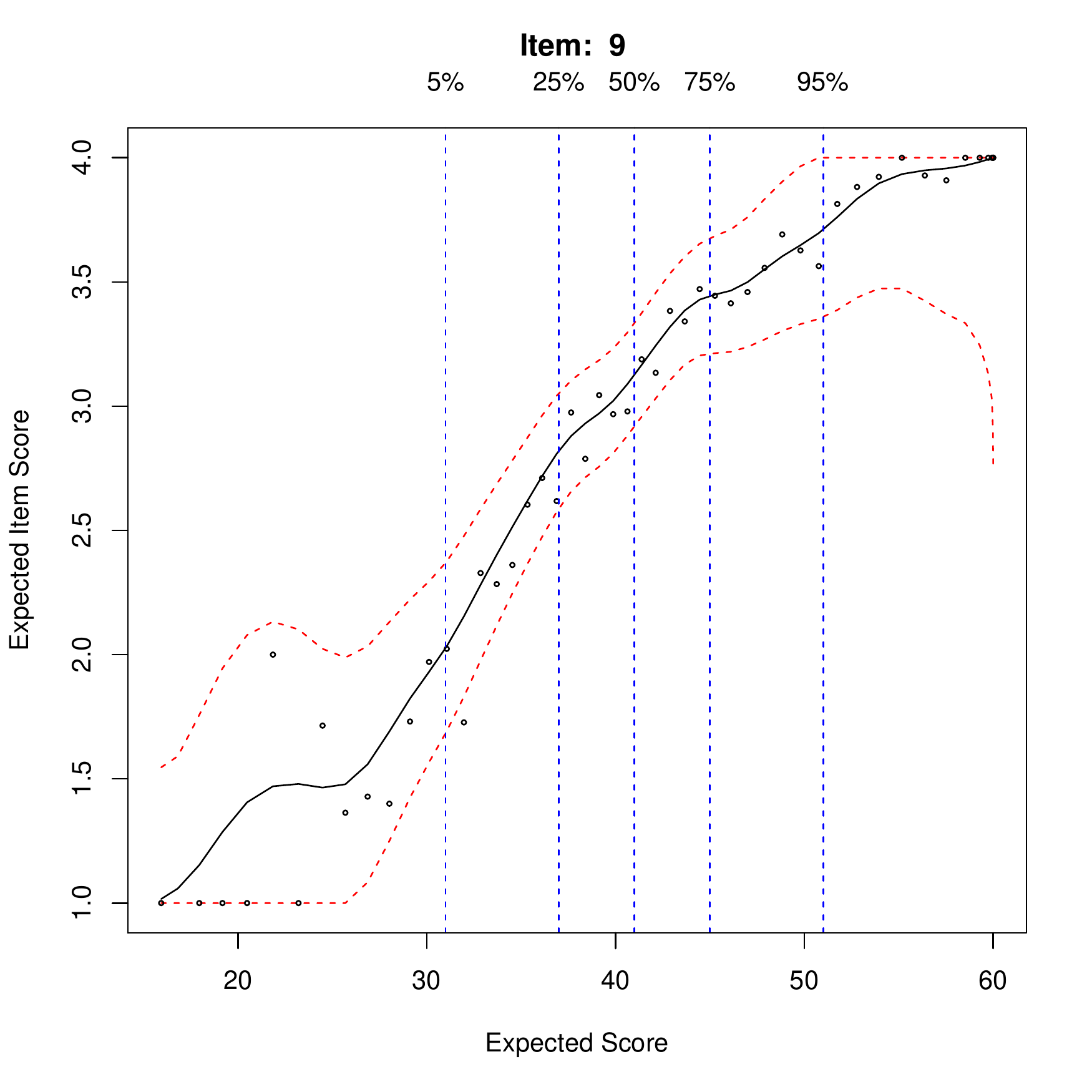}}}
\subfigure[Tetrahedron \label{fig:tetra 9}]
{\resizebox{0.45\textwidth}{!}{\includegraphics{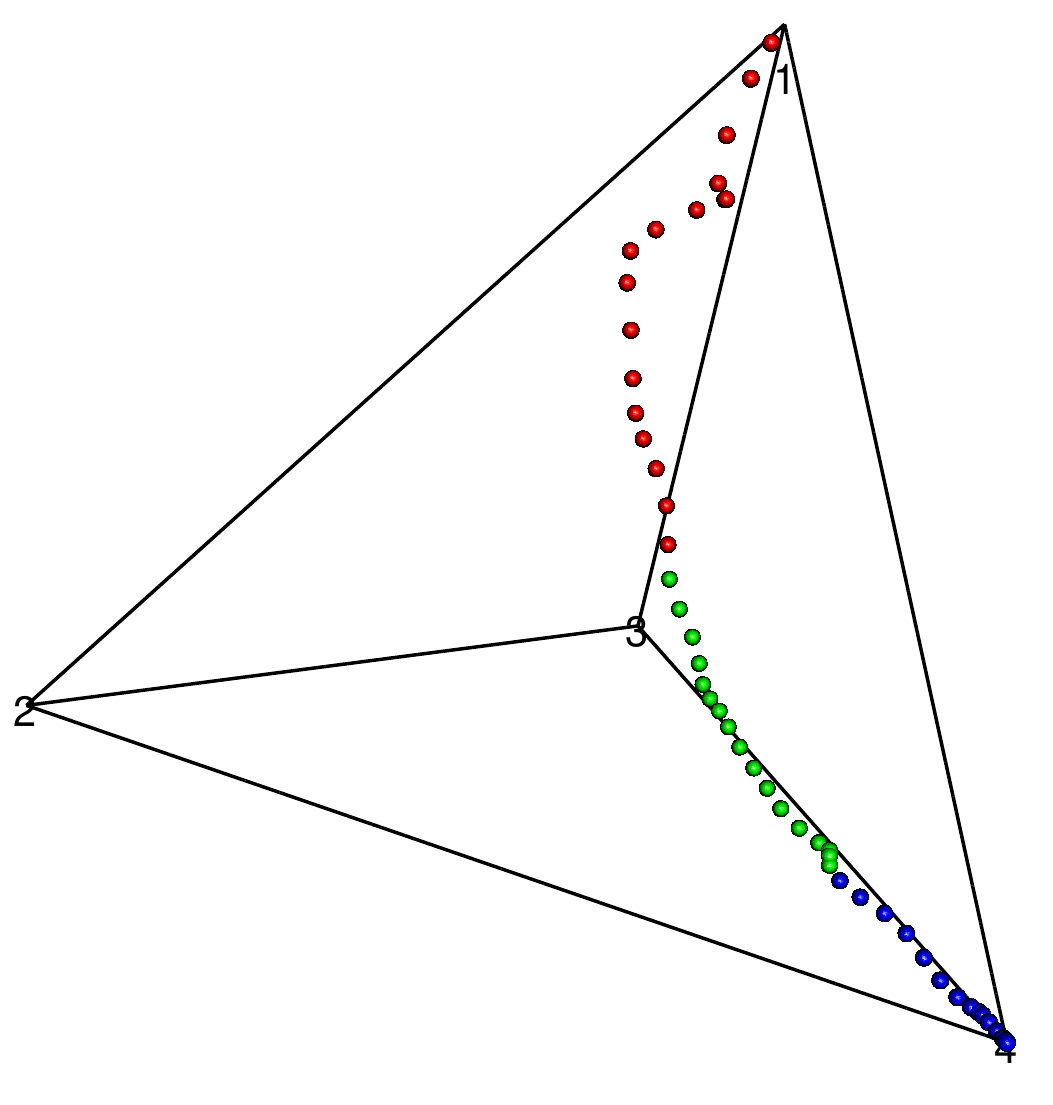}}}
\caption{Item 9 from the voluntary HIV-1 counseling and testing efficacy study group. 
\label{fig:HIV 9}
}
\end{figure}

The code below
\begin{CodeInput}
R> HIVres <- ksIRT(HIV[,-(1:3)], key=HIVkey, format=2, miss="omit")
R> plot(HIVres, plottype="OCC", item=9)
R> plot(HIVres, plottype="EIS", item=9)
R> plot(HIVres, plottype="tetrahedron", item=9) 
\end{CodeInput}
produces the plots,  for $I_9$, displayed in Figure~\ref{fig:HIV 9}.
The option \code{miss="omit"} excludes from the nonparametric analysis all the subjects with at least one omitted answer, leading to a sample of 3473 respondents; the option \code{format=2} specifies that the data contain rating scale items.
Figure~\ref{fig:OCC 9} displays the OCCs for the considered item.
As expected, subjects with the smallest scores are choosing the first option while those with the highest ones are selecting the fourth option.
Generally, as the total scores increase, respondents are approximately estimated to be more likely to choose an higher option and this reflects the typical behavior of a rating scale item.
Figure~\ref{fig:EIS 9} shows the EIS for $I_9$. Note how the expected item score climbs consistently as the total test score increases.
Moreover, the EIS displays a fairly monotone behavior that covers the entire range $\left[1,4\right]$.
Finally, Figure~\ref{fig:tetra 9} shows the tetrahedron for item 9.
It corroborates the good behavior of $I_9$ already seen in Figure~\ref{fig:OCC 9} and Figure~\ref{fig:EIS 9}.
The sequence of points herein, as expected, starts from (the vertex) option 1 and smoothly tends to option 4, passing by option 2 and option 3.

\begin{figure}[!ht]
\centering
\subfigure[Expected scores \label{fig:HIVDIFexp}]
{\resizebox{0.515\textwidth}{!}{\includegraphics{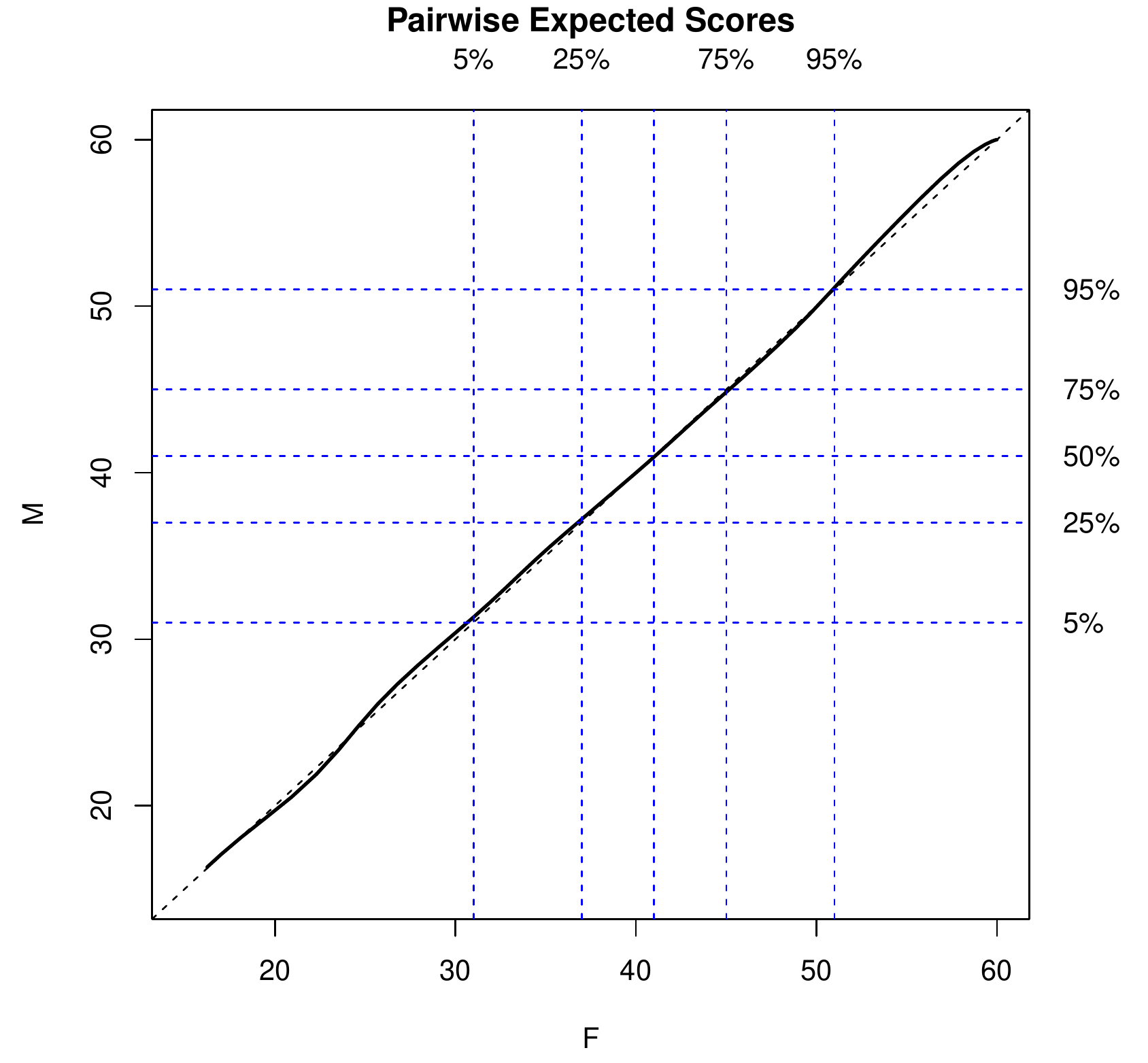}}}
\subfigure[Density \label{fig:HIVDIFden}]
{\resizebox{0.475\textwidth}{!}{\includegraphics{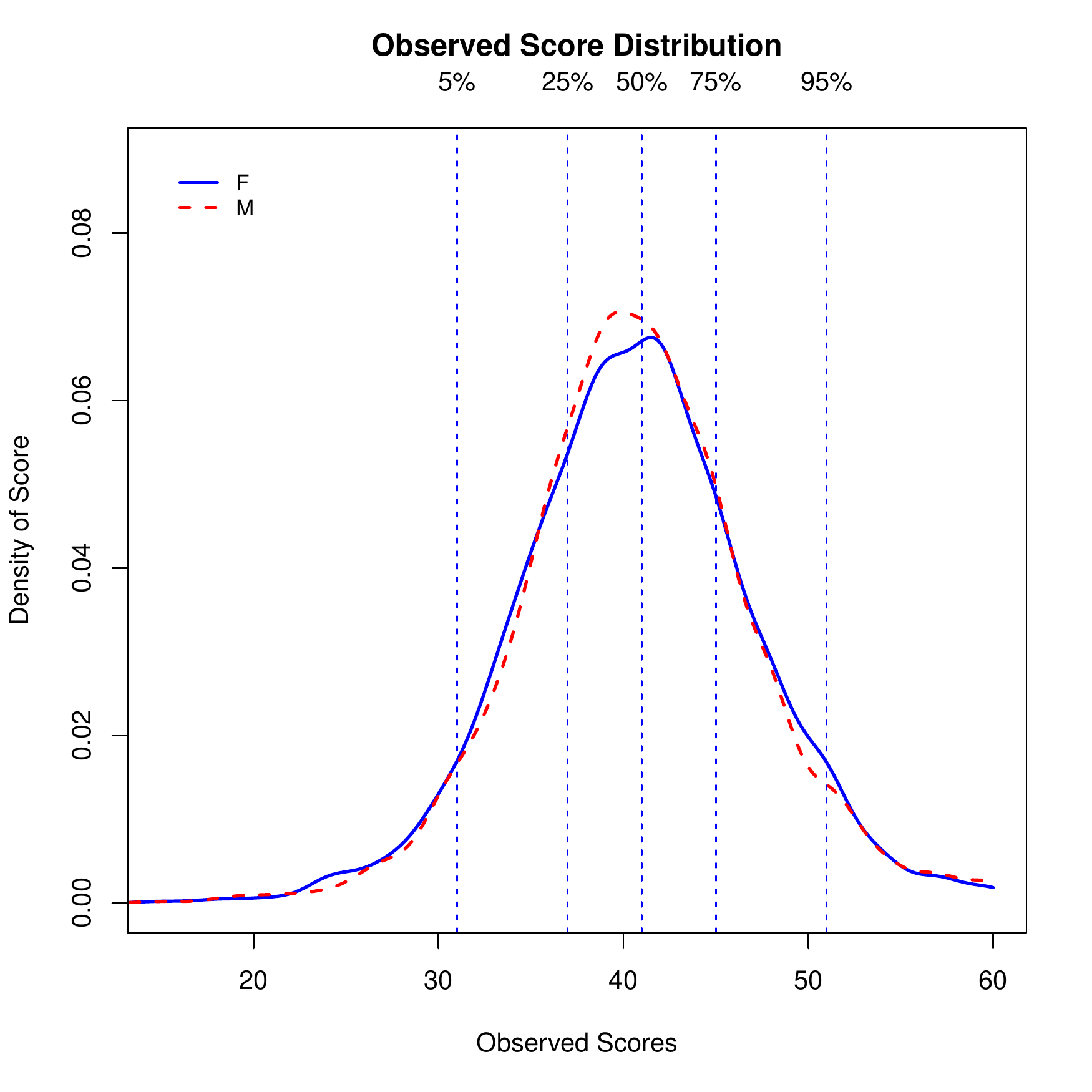}}}
\caption{
Behavior of males (\code{M}) and females (\code{F}) on the test.
In the QQ-plot on the left, the dashed diagonal line indicates the reference situation of no difference in performance for the two groups; the horizontal and vertical dashed blue lines indicate the 5\%, 25\%, 50\%, 75\%, and 95\% quantiles for the two groups. 
}
\label{fig:DIFGENDER}
\end{figure}

The following example demonstrates DIF analysis using the person factor \code{GENDER}. 
To perform this analysis, a new \code{ksIRT} object must be created with the addition of the \code{groups} argument by which the different subgroups may be specified.
In particular, the code
\begin{CodeInput}
R> gr1 <- as.character(HIV$GENDER)
R> DIF1 <- ksIRT(res=HIV[,-(1:3)], key=HIVkey, format=2, groups=gr1,  
 +               miss="omit")
R> plot(DIF1, plottype="expectedDIF", lwd=2)
R> plot(DIF1, plottype="densityDIF", lwd=2)
\end{CodeInput}
produces the plots in Figure~\ref{fig:DIFGENDER}.
Figure~\ref{fig:HIVDIFexp} displays the QQ-plot between the distributions of the expected scores for males and females; if the performances of the two groups are about the same, the relationship will appear as a nearly diagonal line (a dotted diagonal line is plotted as a reference).
Figure~\ref{fig:HIVDIFden} shows the density functions for the two groups.  
Both plots confirm that there is a strong agreement in behavior for males and females with respect to the test.

\begin{figure}[!ht]
\centering
\subfigure[Option 1\label{fig:HIVDIFOCC31}]
{\resizebox{0.49\textwidth}{!}{\includegraphics{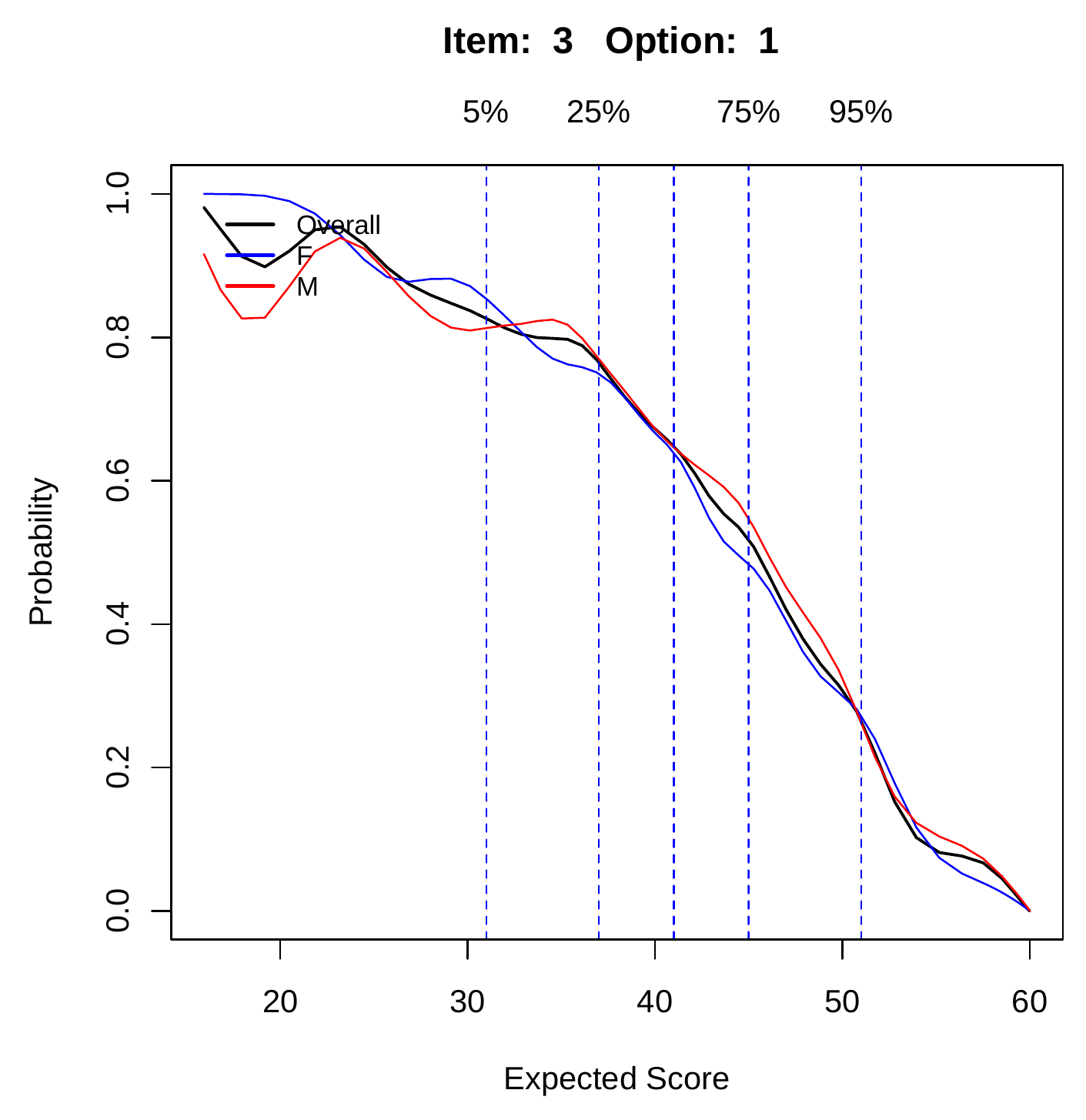}}}
\subfigure[Option 2\label{fig:HIVDIFOCC32}]
{\resizebox{0.49\textwidth}{!}{\includegraphics{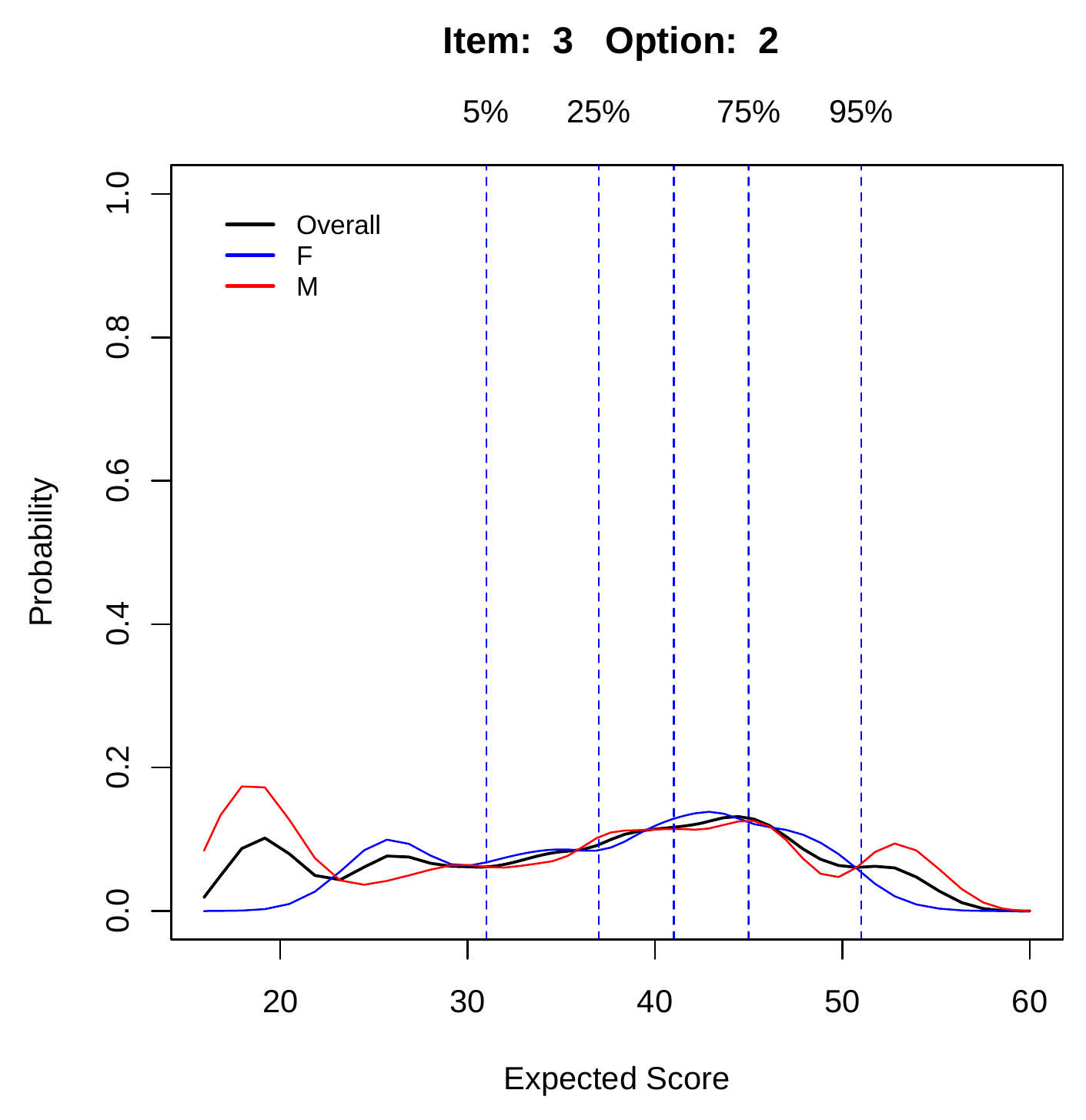}}}
\subfigure[Option 3\label{fig:HIVDIFOCC33}]
{\resizebox{0.49\textwidth}{!}{\includegraphics{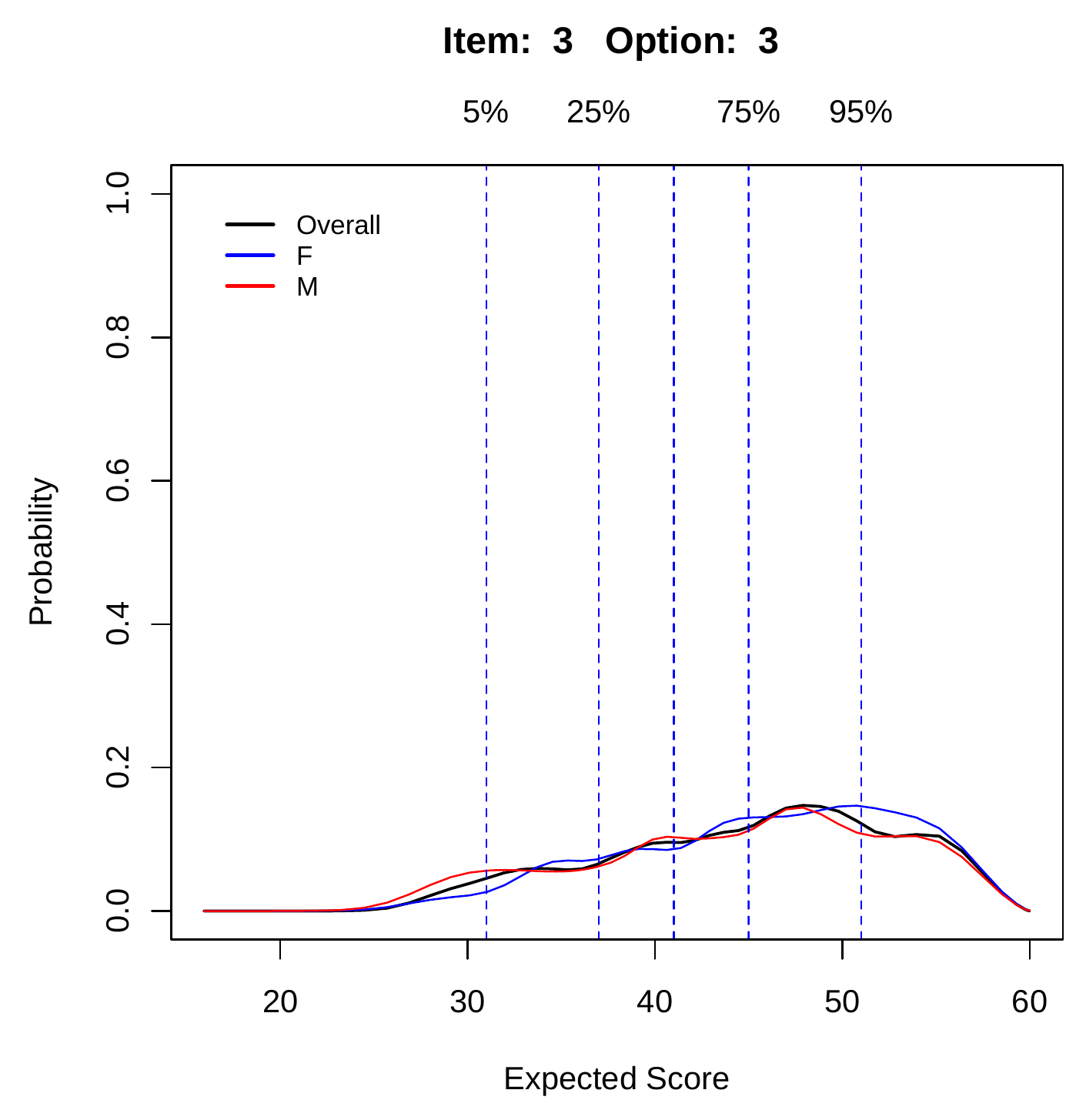}}}
\subfigure[Option 4\label{fig:HIVDIFOCC34}]
{\resizebox{0.49\textwidth}{!}{\includegraphics{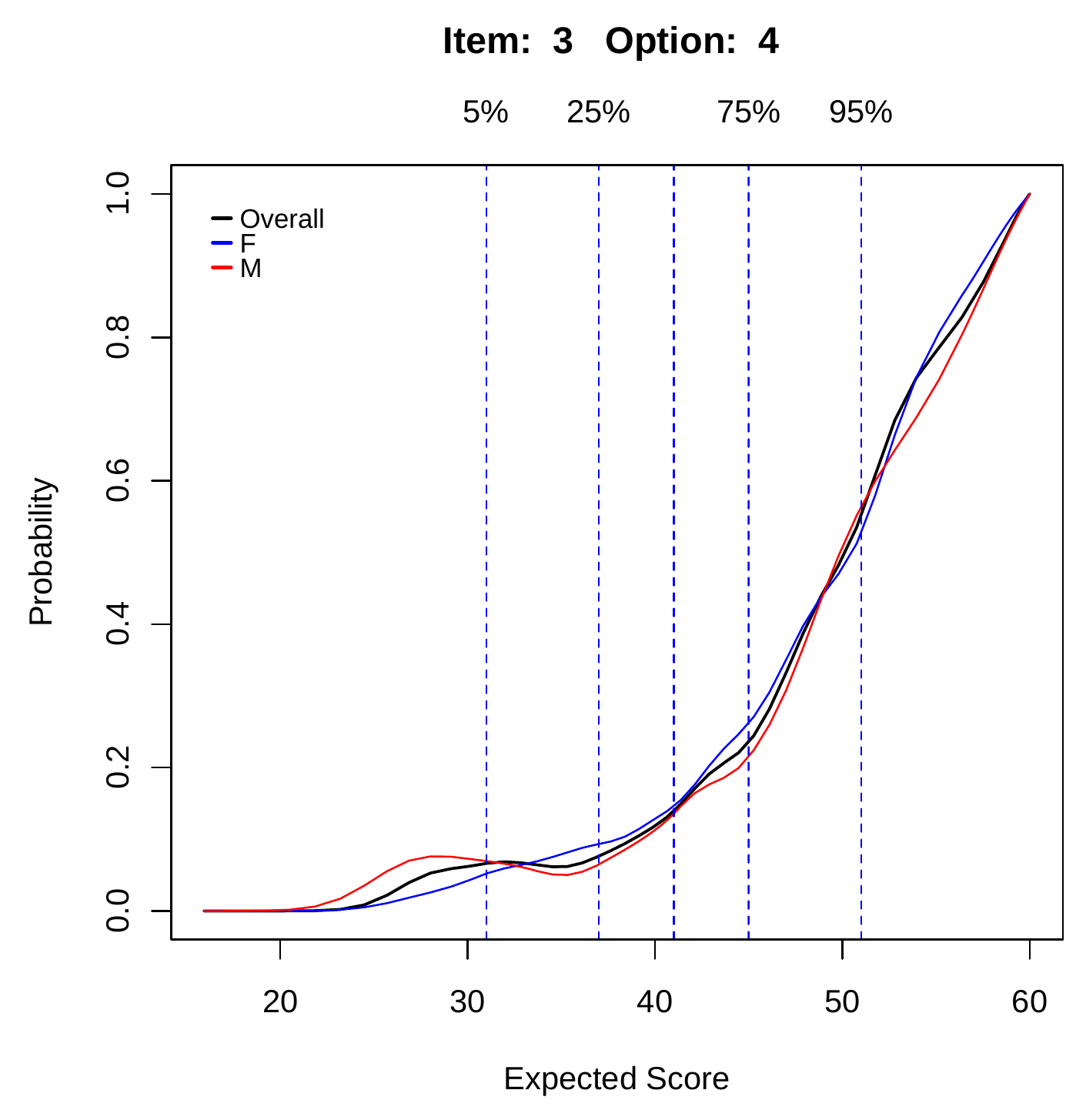}}}
\caption{
OCCs, for males and females, related to item 3 of the voluntary HIV-1 counseling and testing efficacy study group. 
The overall OCCs are superimposed.
}
\label{fig:HIVDIFOCC3}
\end{figure}

\begin{figure}[!ht]
\centering
\resizebox{0.68\textwidth}{!}{
\includegraphics{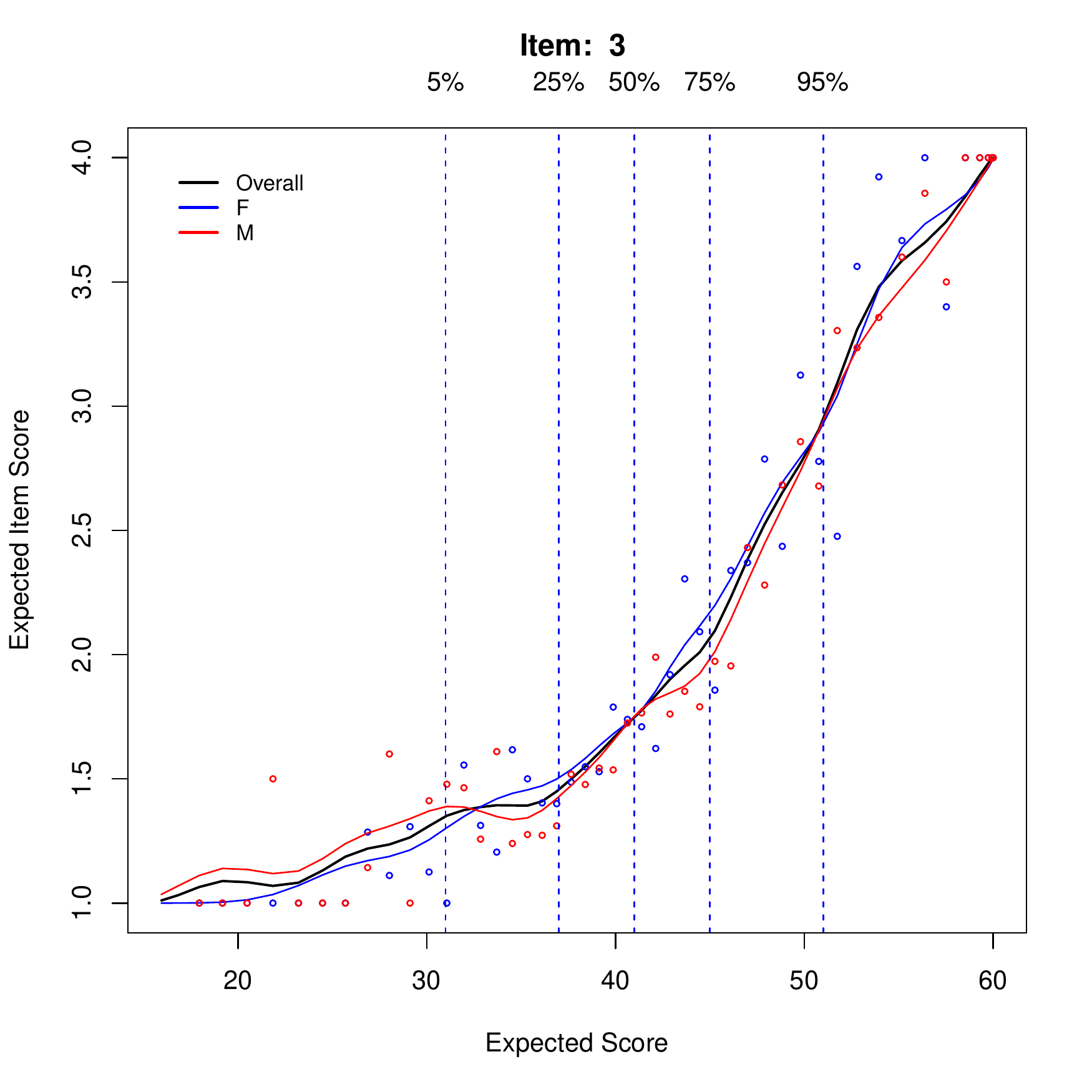}
}
\caption{
EIS, of males and females, for item 3. 
}
\label{fig:HIVDIFEIS3} 
\end{figure}

After this preliminary phase, the DIF analysis proceeds by considering the item by item group comparisons.
Figure~\ref{fig:HIVDIFOCC3}, obtained via the command
\begin{CodeInput}
R> plot(DIF1, plottype="OCCDIF", cex=0.5, item=3)
\end{CodeInput}
displays the OCCs for the (rating scale) item $I_3$.     
These plots allow the user to compare the two groups at the item level. 
Lack of DIF is evident by nearly overlapping OCCs for all the four options.
DIF may also be evaluated in terms of the expected score of the groups, as displayed in Figure~\ref{fig:HIVDIFEIS3}.
This plot is obtained with the code
\begin{CodeInput}
R> plot(DIF1, plottype="EISDIF", cex=0.5, item=3)
\end{CodeInput}
The different color points on the plot represent how individuals from the groups actually scored on the item.
Although we have focused the attention only on $I_3$, similar results are obtained for all of the other items in $\mathcal{I}$, and this confirms as \code{GENDER} is not a variable producing DIF in this study.
This result is corroborated in \citet{Bert:Musc:Punz:Item:2010}. 
Note that, for both OCCs and EISs, it is possible to add confidence intervals through the \code{alpha} argument.

\begin{figure}[!ht]
\centering
\subfigure[QQ-plot (\code{Ken} vs. \code{Tan}) \label{fig:ExpTanKen}]
{\resizebox{0.49\textwidth}{!}{\includegraphics{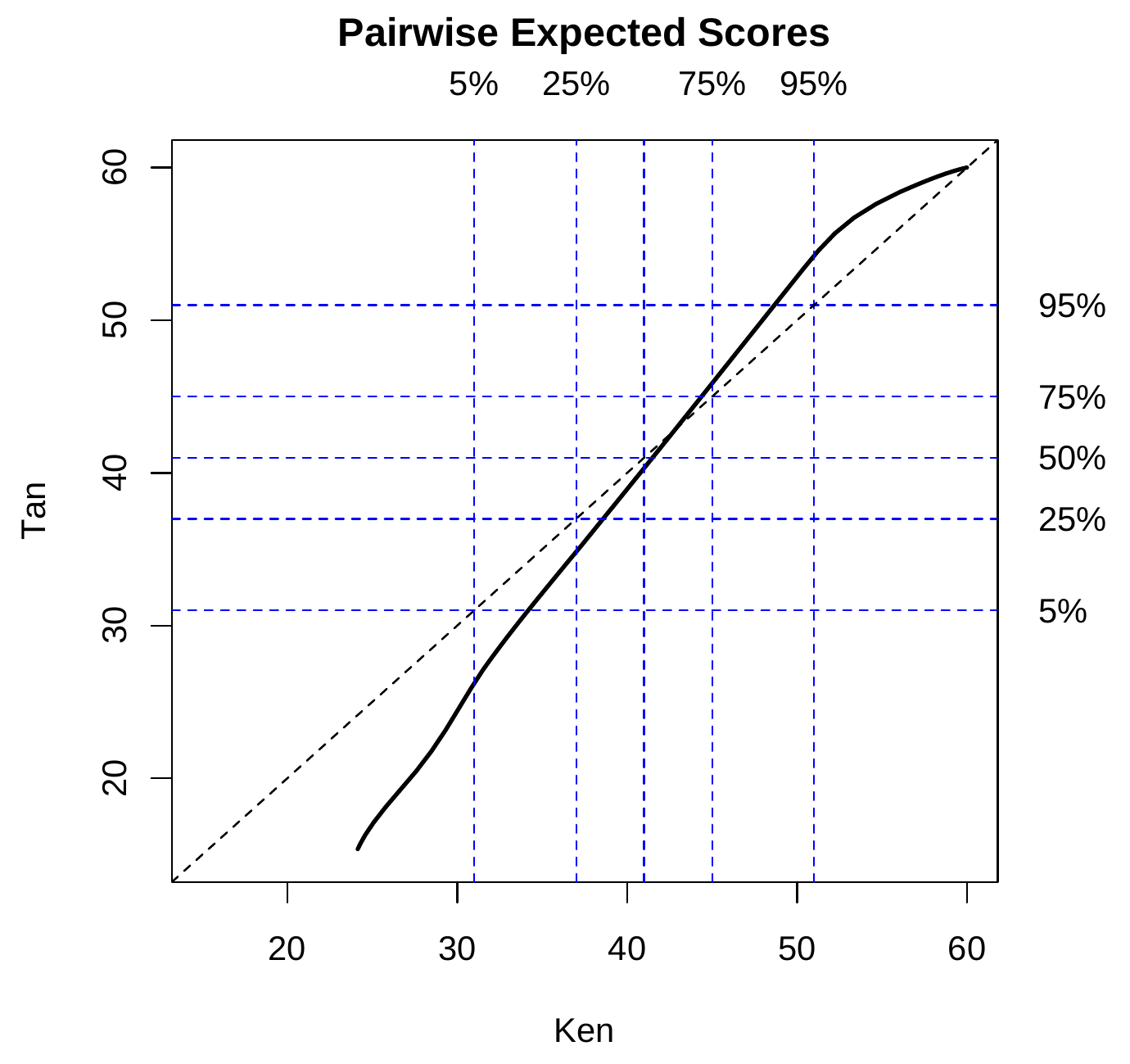}}}
\subfigure[QQ-plot (\code{Tan} vs. \code{Tri}) \label{fig:ExpTriTan}]
{\resizebox{0.49\textwidth}{!}{\includegraphics{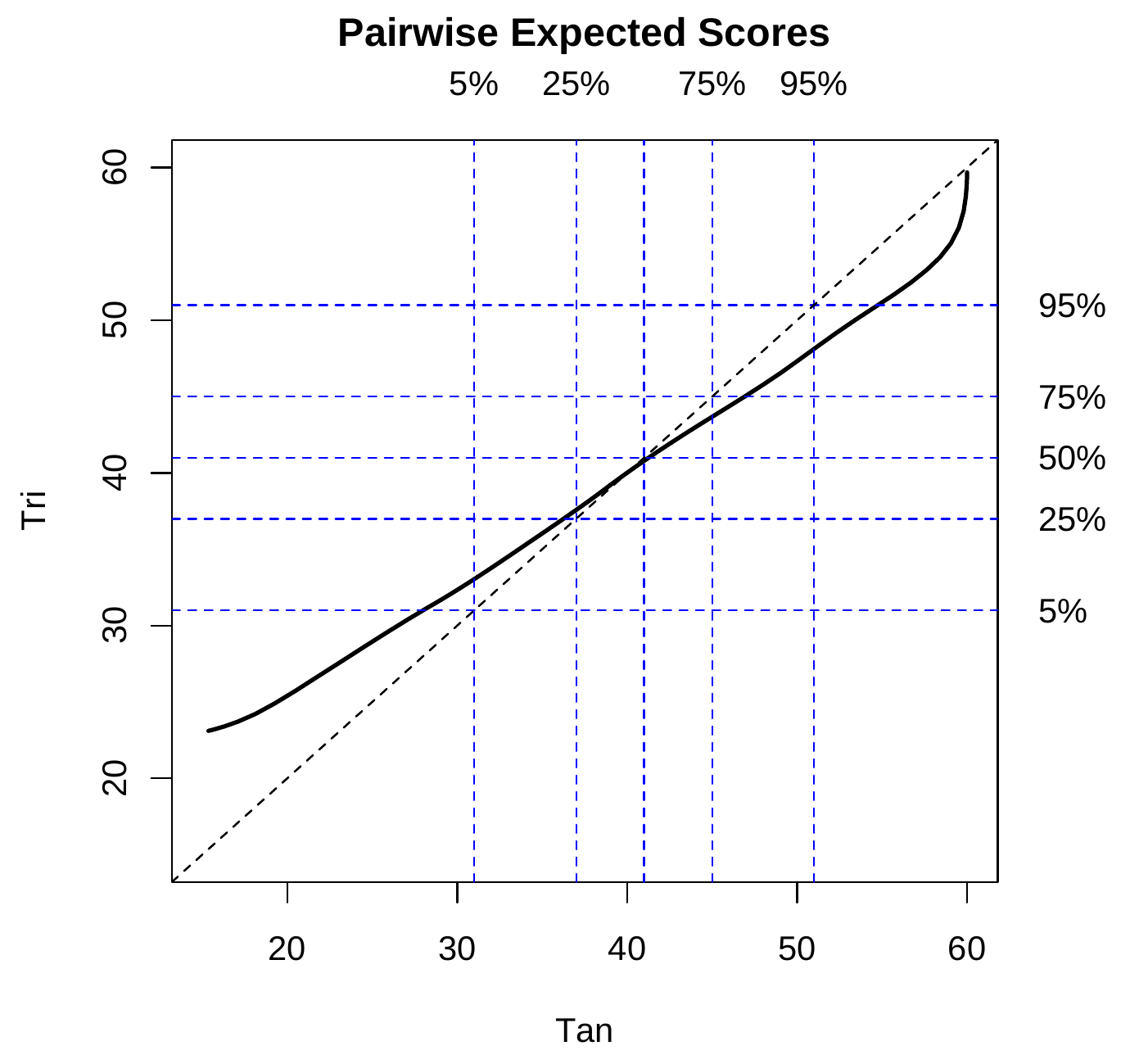}}}
\subfigure[QQ-plot (\code{Ken} vs. \code{Tri}) \label{fig:ExpTriKen}]
{\resizebox{0.49\textwidth}{!}{\includegraphics{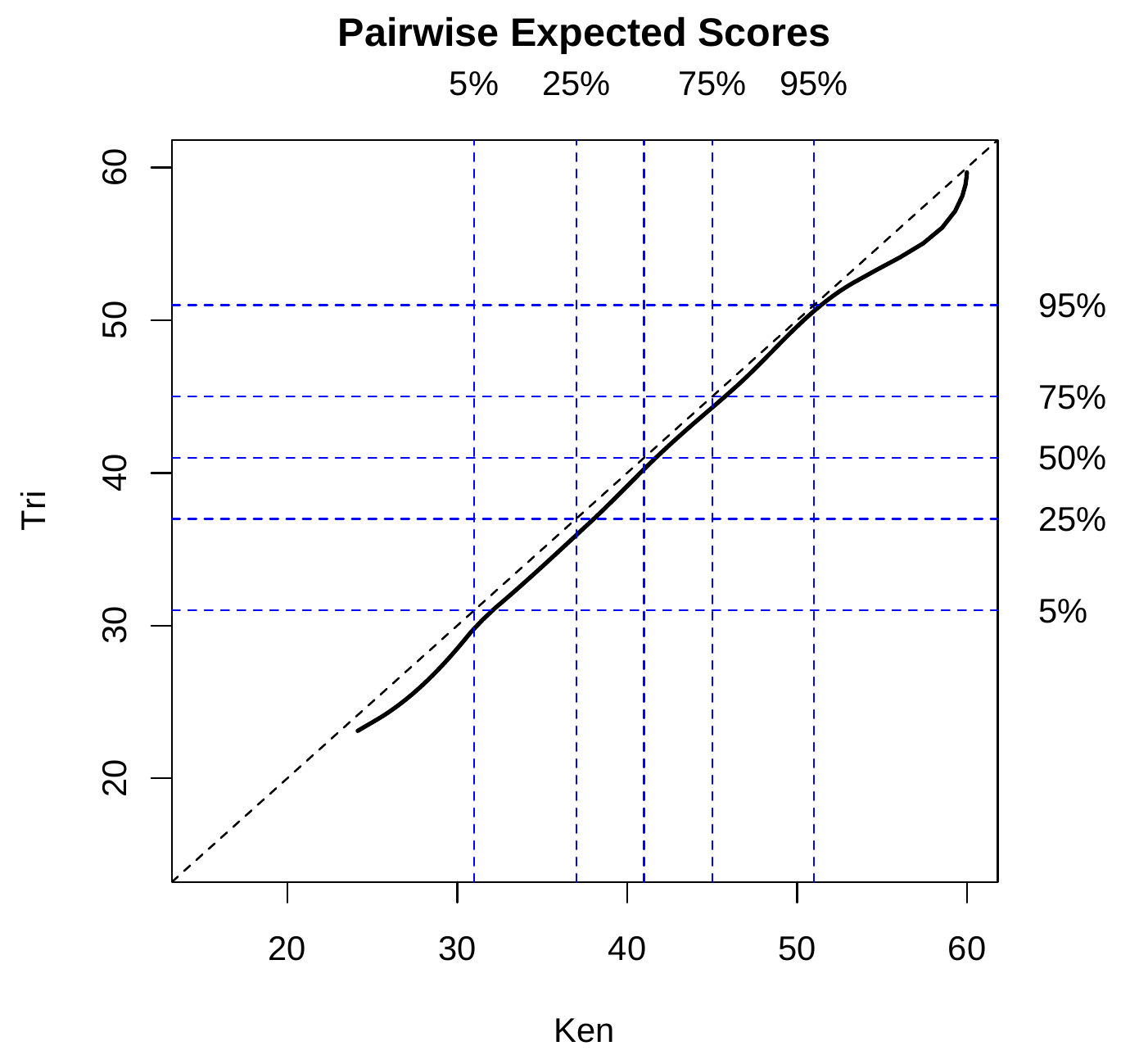}}}
\subfigure[Total score density \label{fig:HIVDIFSden}]
{\resizebox{0.49\textwidth}{!}{\includegraphics{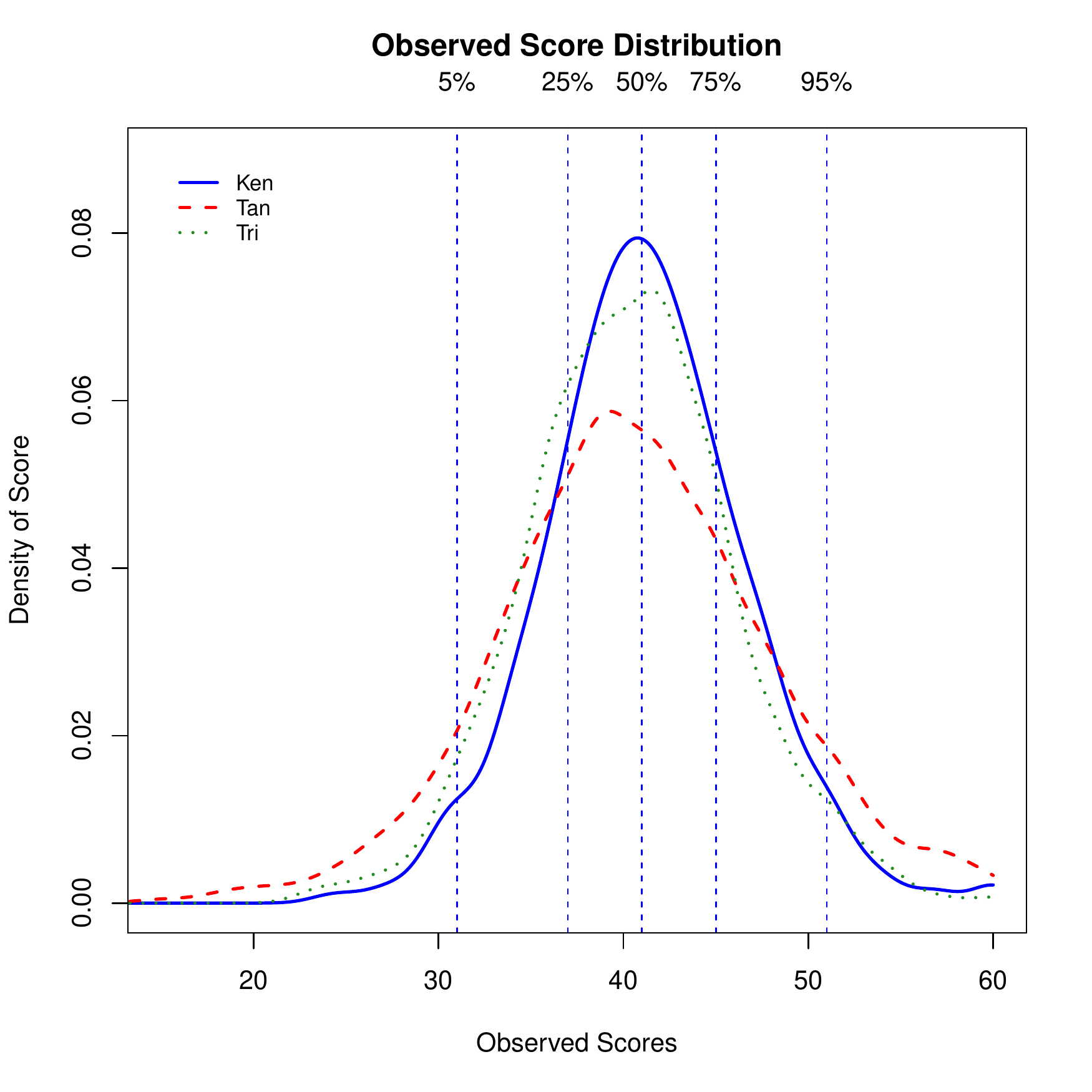}}}
\caption{
Behavior of people from Kenya (\code{Ken}), Tanzania (\code{Tan}), and Trinidad (\code{Tri}), on the test.
In all the pairwise QQ-plots, the dashed diagonal line indicates the reference situation of no difference in performance for the two groups; the horizontal and vertical dashed blue lines indicate the 5\%, 25\%, 50\%, 75\%, and 95\% quantiles for the two groups. 
}
\label{fig:DIFSITE}
\end{figure}

\begin{figure}[!ht]
\centering
\subfigure[Item 6 \label{fig:HIVDIFEISSITE6}]
{\resizebox{0.49\textwidth}{!}{\includegraphics{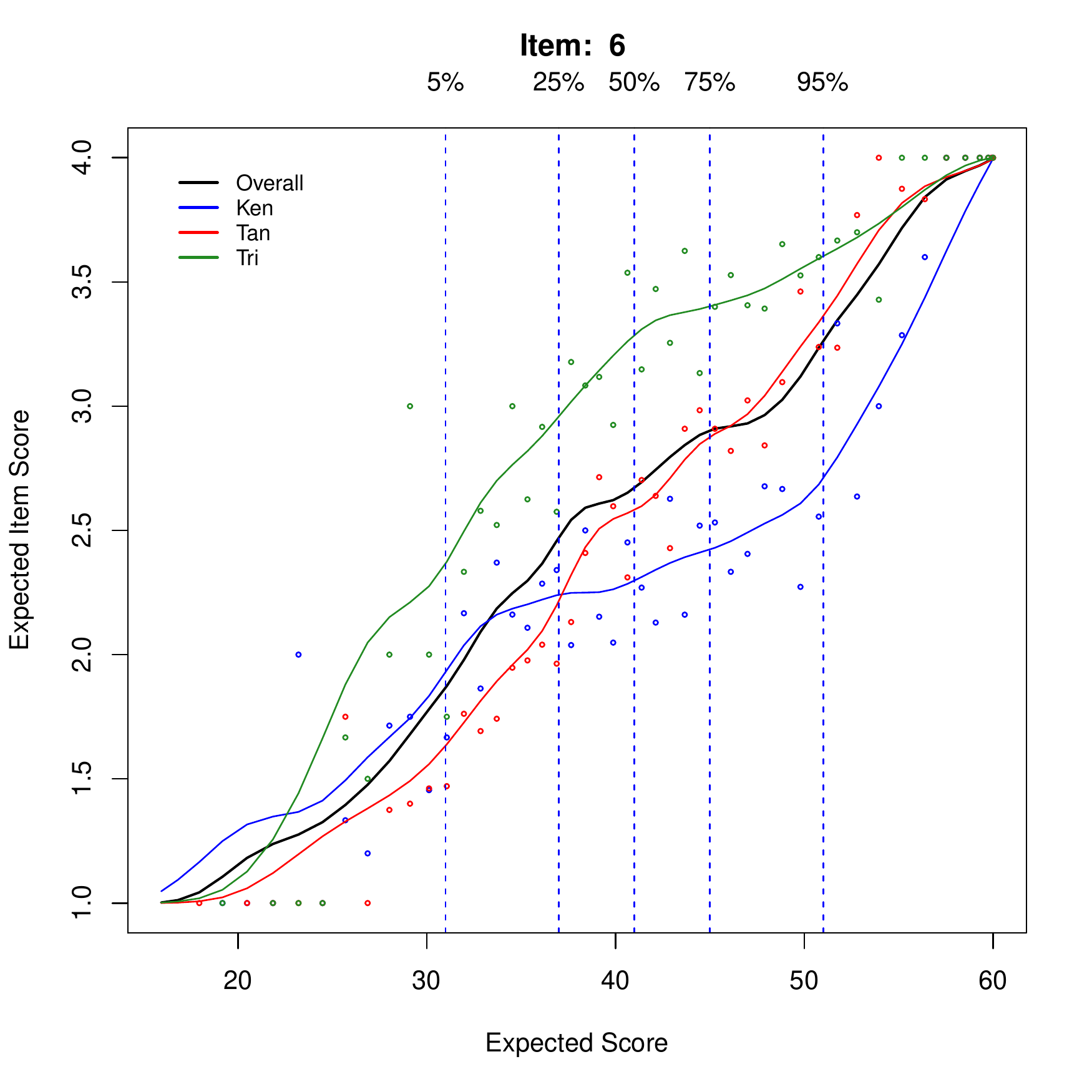}}}
\subfigure[Item 11 \label{fig:HIVDIFEISSITE11}]
{\resizebox{0.49\textwidth}{!}{\includegraphics{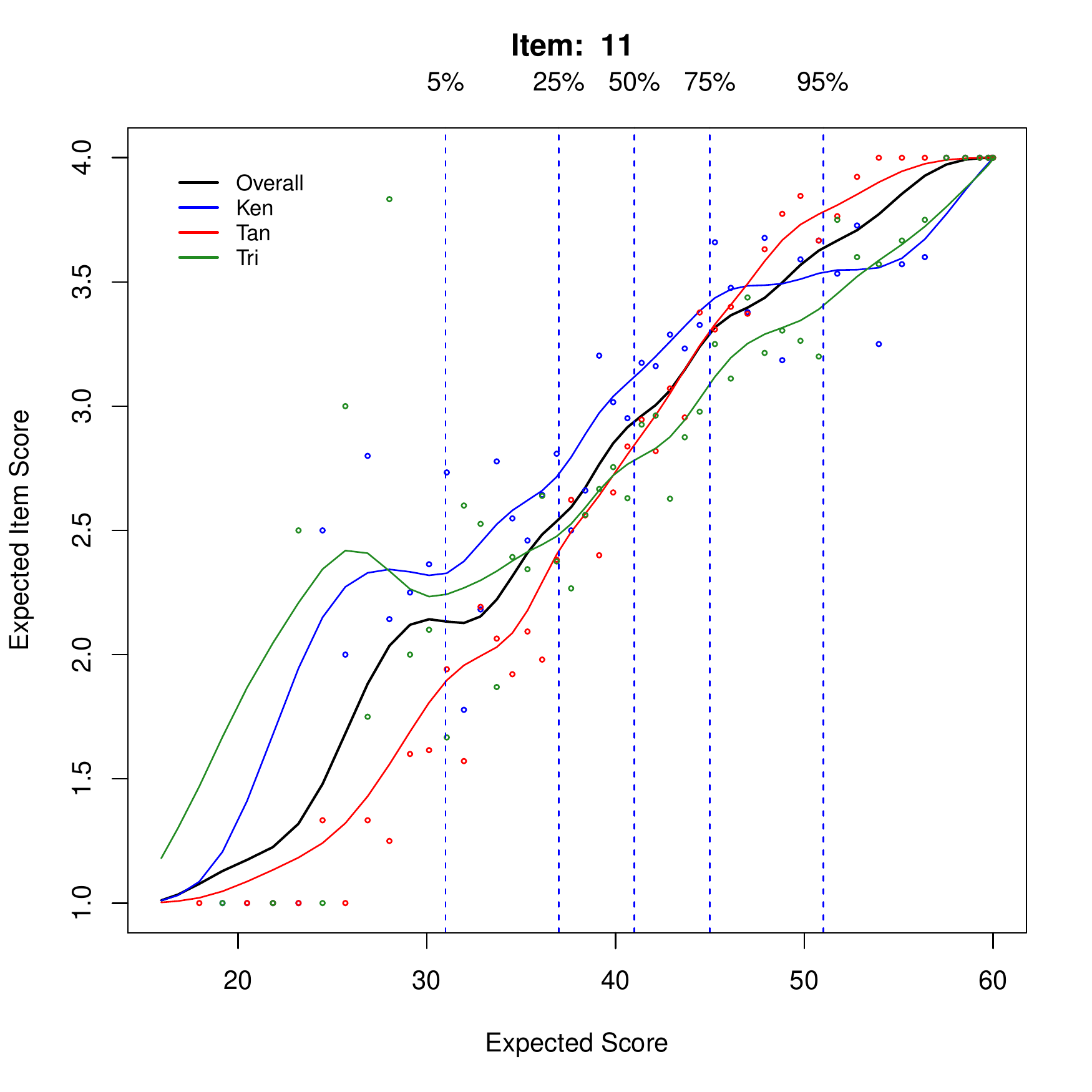}}}
\caption{
EISs, for people from Kenya (\code{Ken}), Tanzania (\code{Tan}), and Trinidad (\code{Tri}), for items 6 and 11 of the voluntary HIV-1 counseling and testing efficacy study group. 
}
\label{fig:HIVDIFEISSITE}
\end{figure}

The code
\begin{CodeInput}
R> gr2 <- as.character(HIV$SITE)
R> DIF2 <- ksIRT(res=HIV[,-(1:3)], key=HIVkey, format=2, groups=gr2,
 +               miss="omit")
R> plot(DIF2, plottype="expectedDIF", lwd=2)
R> plot(DIF2, plottype="densityDIF", lwd=2)
\end{CodeInput} 
produces separated plots for subjects with different \code{SITE} levels (Figure~\ref{fig:DIFSITE}).
Among the 3473 subjects answering to all the 15 items, 984 come from Trinidad, 1143 from Kenya and 1346 from Tanzania.
As highlighted by \citet{Bert:Musc:Punz:Item:2010}, there are differences among these groups, and Figure~\ref{fig:DIFSITE} shows this.
The three pairwise QQ-plots of the expected score distributions show that there is a slight dominance of people from Kenia over people from Trinidad (in the sense that people from Kenia have, in distribution, a slightly more positive attitude toward condom use than people from Trinidad), and a large discrepancy between the performances of people from Tanzania and those of the other two groups, as shown in Figure~\ref{fig:ExpTanKen} and Figure~\ref{fig:ExpTriTan}.
The above dominance, and the peculiar behavior of people from Tanzania compared with the other countries, can be also noted by looking at the observed total score densities in Figure~\ref{fig:HIVDIFSden}.
Here, there is higher variability in the total score for people from Tanzania.
But what about DIF? 
The command 
\begin{CodeInput}
R> plot(DIF2, plottype="EISDIF", item=c(6,11))
\end{CodeInput}
produces, for $I_6$ and $I_{11}$, the EISs in Figure~\ref{fig:HIVDIFEISSITE}.
In both the plots we have a graphical indication of the presence of DIF, and this confirms the results by \citet{Bert:Musc:Punz:Item:2010} that detect \code{SITE}-based DIF for these and other items in the test.

\section[Conclusions]{Conclusions}
\label{sec:conclusions}

In this paper, package \pkg{KernSmoothIRT} for the \proglang{R} environment, which allows for kernel smoothing within the IRT context, has been introduced. 
Two applications have been discussed, along with some theoretical and practical issues.

The advantages of nonparametric IRT modeling  are well known. 
\citet{Rams:test:2000} recommends its application, at least as an exploratory tool, to guide users over the choice of an appropriate parametric model. 
Moreover, while currently most IRT analyses are conducted  with parametric models, quite often the assumptions underling parametric IRT modeling are not preliminarily checked.
One reason for this may be the lack, apart from \proglang{TestGraf}, of available software.  
\proglang{TestGraf} has set a milestone on this field as the first computer program to implement a kernel smoothing approach to IRT and has been the prominent software for years. 
Compared to \proglang{TestGraf}, \pkg{KernSmoothIRT} has the major advantage of running within the \proglang{R} environment. 
Users do not have to export their results into another piece of software in order to perform non-standard data analysis, to produce customized plots or to perform parametric IRT using any of several packages available in \proglang{R}. 
Furthermore, \pkg{KernSmoothIRT} allows more flexibility in bandwidth and kernel selection, as well as in handling missing values.

We believe that \pkg{KernSmoothIRT} may prove useful to educators, psychologists, and other researchers developing questionnaires, enabling them to spot ill-posed questions and to formulate more plausible wrong options.
Future works will consider extending the package by allowing for kernel smoothing estimation of test and item information functions. 
Although well-established in parametric IRT, information functions present serious statistical problems in NIRT context, as underlined by \citet[][p.~66]{Rams:test:2000}. 
Currently available nonparametric-based IRT programs, such as \proglang{TestGraf}, estimate test and item information functions based on parametric OCCs.

\section*{Acknowledgments}

The Voluntary HIV1 Counseling and Testing study was sponsored by UNAIDS/WHO, AIDSCAP/Family Health International, and the Center for AIDS Prevention Studies at the University of California, San Francisco.

\bibliography{arXiv}

\end{document}